\documentclass[10pt, aps, prx, superscriptaddress, twocolumn]{revtex4-2}
\usepackage[version=3]{mhchem}
\usepackage{amsmath, amssymb, amsfonts}
\usepackage{graphicx}
\usepackage{float}
\usepackage{xcolor}
\usepackage{bm}
\usepackage{braket}
\usepackage{siunitx}  
\usepackage{ulem}
\usepackage[colorlinks=true, allcolors=blue]{hyperref}

\newif\ifshowmarkup
\showmarkupfalse
\ifshowmarkup
    
    \newcommand{\var}[1]{\textsf{\color[HTML]{4A6FA5}{(variant: #1)}}}
    \newcommand{\old}[1]{\textsf{\color{gray}(old: #1)}}
    \newcommand{\aaron}[1]{\textsf{\color[HTML]{6E3CBC}{(AS: #1)}}}
    \newcommand{\aviram}[1]{\textsf{\color[HTML]{6E7F10}{(AU: #1)}}}
    \newcommand{\xia}[1]{\textsf{\color[HTML]{4CAF50}{(LQX: #1)}}}
    \newcommand{\YK}[1]{\textsf{\color[HTML]{4CAF50}{(YK: #1)}}}
\else
    \renewcommand{\sout}[1]{}     
          
    \newcommand{\var}[1]{}        
    \newcommand{\old}[1]{}        
    \newcommand{\aaron}[1]{}   
    \newcommand{\aviram}[1]{} 
    \newcommand{\xia}[1]{} 
    \newcommand{\YK}[1]{} 
\fi

\newcommand{\addcite}[1]{\textsf{\color{red!80!black}[\#]}}

\begin{document}

\title{Quantized Transport through a Supermoir\'e Chern Mosaic}

\def\StanfordPhys{Department of Physics, Stanford University, Stanford, CA 94305}
\def\SIMES{Stanford Institute for Materials and Energy Sciences, SLAC National Accelerator Laboratory, Menlo Park, CA 94025}
\def\StanfordMatsci{Department of Materials Science and Engineering, Stanford University, Stanford, CA 94305}

\def\MITPhys{Department of Physics, Massachusetts Institute of Technology, Cambridge, Massachusetts 02139, USA}

\def\UTDPhys{Department of Physics, University of Texas at Dallas, Richardson, Texas 75080, USA.}

\author{Li-Qiao~Xia}
\thanks{These authors contributed equally to this work.}
\affiliation{\MITPhys}

\author{Aviram~Uri}
\thanks{These authors contributed equally to this work.}
\affiliation{\MITPhys}

\author{Zachary~W.~Gomez}
\affiliation{\StanfordPhys}
\affiliation{\SIMES}

\author{Molly~P.~Andersen}
\affiliation{\StanfordMatsci}
\affiliation{\SIMES}

\author{Julian~May-Mann}
\affiliation{\StanfordPhys}

\author{Kenji~Watanabe}
\affiliation{Research Center for Electronic and Optical Materials, National Institute for Materials Science, 1-1 Namiki, Tsukuba 305-0044, Japan}

\author{Takashi~Taniguchi}
\affiliation{Research Center for Materials Nanoarchitectonics, National Institute for Materials Science,  1-1 Namiki, Tsukuba 305-0044, Japan}

\author{Trithep~Devakul}
\affiliation{\StanfordPhys}

\author{Yves~H.~Kwan}
\email{yveshon.kwan@utdallas.edu}
\affiliation{\UTDPhys}

\author{Pablo~Jarillo-Herrero}
\email{pjarillo@mit.edu}
\affiliation{\MITPhys}

\author{Aaron~Sharpe}
\email{aaron.sharpe@stanford.edu}
\affiliation{\StanfordPhys}
\affiliation{\SIMES}

\date{\today}

\begin{abstract}

Magic-angle helical trilayer graphene---three graphene layers sequentially twisted in the same direction by $\sim1.8^\circ$---relaxes into a mosaic of domains that, at zero field, carry opposite valley-resolved Chern numbers, with boundaries hosting a network of gapless conducting modes.
Charge transport through this network depends sensitively on how the modes connect and scatter, making well-quantized transport unlikely.
Contrary to this expectation, we observe a field-induced Chern gap with Chern number $C=-6$ emanating from charge neutrality; in this gap, the Hall resistance is quantized to within $2\%$ of the expected value, $-h/6e^2$, at \SI{4.6}{K}.
We explain this behavior using both Hofstadter and orbital Zeeman calculations, which show that a moderate magnetic field drives a valley-selective topological transition.
Above the transition, the total Chern number of the occupied states in each spin-valley flavor becomes identical across neighboring domains, and the domain-wall modes can become gapped.
Though the central valence-band Chern numbers still differ between the two domain types, the observed quantized transport attests to a global gap.
\end{abstract}

\maketitle

%%%%%%%%%%%%%%%%%%%%%%%%%%%%%%%%%%%%%%%%%%%%%%%%%%%%%%%%%
\section*{Introduction}\label{sec:main_fci}

Two-dimensional insulating phases of matter are characterized by an integer topological invariant, the Chern number, $C$. 
An interface where the Chern number changes hosts topologically protected gapless modes, the number of which is set by the change in $C$.
For a sample with $C\neq0$, such a boundary occurs naturally at its physical edge, where the adjoining vacuum is topologically trivial ($C=0$).
Consequently, the edges host $|C|$ chiral edge modes, resulting in vanishing longitudinal resistance, $R_{xx}=0$, and quantized Hall resistance, $R_{yx} = h/Ce^2$, where $e$ is the fundamental charge and $h$ is Planck's constant. 

\begin{figure*}[t!]
\centering
\includegraphics{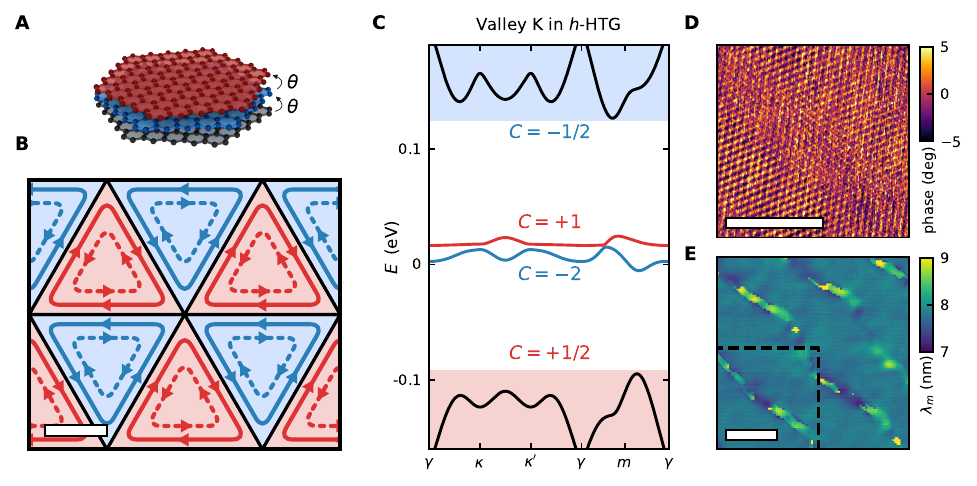}
\caption{\textbf{Helical Trilayer Graphene (HTG).} 
(\textbf{A}) Schematic of HTG. 
(\textbf{B}) Idealized supermoir\'e structure of HTG near the magic angle of $\theta=\SI{1.8}{\degree}$. 
Moir\'e-periodic h-HTG (red shading) and $\bar{\text{h}}$-HTG (blue shading) domains formed by lattice relaxation are indicated. 
When the system is charge-neutral and time-reversal symmetric, each domain wall hosts six pairs of counter-propagating modes, which we schematically draw as solid ($K$) and dashed ($K'$) lines, each representing three modes. 
The lines are spatially separated for clarity.
(\textbf{C}) Single-particle band structure for the $K$ valley of h-HTG  (identical to $K'$ in $\bar{\text{h}}$-HTG).
The Chern numbers of the central bands and the total Chern number of the remote bands are indicated. 
Broken $C_{2z}$ within a domain allows the local bands to carry valley-contrasting Chern numbers.
At $B=0$, within a valley, the band structures of h-HTG and $\bar{\text{h}}$-HTG are related by $C_{2z}\mathcal{T}$. 
(\textbf{D}) Torsional force microscopy image of an HTG sample.
Phase variations reveal a moir\'e with an approximate period of $8$\,nm.
Modulations in the contrast of the phase likely correspond to the supermoir\'e structure.
For the full range of data, see Fig.~\ref{sup_fig:full_tfm}.
(\textbf{E}) Local moir\'e period extracted from $\SI{75}{nm} \times \SI{75}{nm}$ windowed FFTs of the full TFM scan. 
See Sec.~\ref{sup_sec:TFM} for details of this calculation. 
Domain walls are identified as an increase or decrease in the local moir\'e period relative to the predominant period.
Black dashed box indicates the region shown in (\textbf{D}). 
Strain can dramatically distort the supermoir\'e structure (see Sec.~\ref{sup_sec:twist}).
All scale bars are $\SI{100}{nm}$.
}
\label{fig:schematic}
\end{figure*}

Helical trilayer graphene (HTG) consists of three graphene layers sequentially twisted in the same direction by the same angle $\theta$ (Fig.~\ref{fig:schematic}A)~\cite{devakul_magic-angle_2023, yang_multi-moire_2024,
kwan2024strong, hoke_imaging_2024, xia_topological_2025,
bocarsly2026reversals}.
The two superimposed moir\'e patterns yield a higher-order supermoir\'e modulation that further tunes the electronic structure on a larger length scale~\cite{zhu_twisted_2020, popov_magic_2023, nakatsuji2023multiscale, yang_multi-moire_2024, foo_extended_2024, hesp_cryogenic_2024, xia_topological_2025, xie_strong_2025, Zhou2026supermoire}.
Near the magic angle of $\theta\simeq1.8^\circ$, HTG exhibits a rich phase diagram of strongly correlated states, including states with an anomalous Hall effect (AHE)~\cite{xia_topological_2025}.

Lattice relaxation favors regions where $xy$-inversion ($C_{2z}$) symmetry is locally broken~\cite{hoke_imaging_2024}, producing two types of domains that tile the sample so that $C_{2z}$ is preserved globally~\cite{devakul_magic-angle_2023}. 
When time-reversal symmetry is unbroken, the two domain types carry opposite valley-resolved Chern numbers.
In the presence of conserved spin-valley quantum numbers (hereafter referred to as flavors), the change in Chern number within each flavor determines the number of flavor-resolved gapless modes at an interface, even if the total Chern number does not change.
These domains produce a network of edge modes at domain walls, which remain topologically protected if intervalley scattering is negligible (Fig.~\ref{fig:schematic}B). 
For such a network, resistance measurements depend sensitively on the position of the voltage probes and the details of the equilibration between edge modes~\cite{Sayak, rickhaus_transport_2018, williamsQuantumHallEffect2007, abaninChargeSpinTransport2007}.
Prior observations of a large longitudinal resistance and unquantized AHE within HTG are consistent with such a network~\cite{xia_topological_2025}.

Here, in contrast to this expectation, we show that under a perpendicular magnetic field, a robust insulating state develops, with $R_{yx}=-h/6e^2$ to within $2\%$ and longitudinal resistance $R_{xx}\lesssim \SI{100}{\Omega}$. 
This state emanates from the charge-neutrality point (CNP) and follows the  St\v{r}eda relation $(h/e)(\partial n/\partial B)=C=-6$ expected for this value of the Hall resistance, a behavior characteristic of a global Chern insulator, not a network of edge modes.
For the internal domain walls to become gapped, neighboring domains must have the same Chern number for each conserved flavor. 
We demonstrate that this change from a Chern mosaic to a global Chern-insulating state arises from a field-induced, valley-selective topological transition.

% ---------------------------------------------------------------------------
% ---------------------------------------------------------------------------

%%%%%%%%%%%%%%%%%%%%%%%%%%%%%%%%%%%%%%%%%%%%%%%%%%%%%%%%%
\section*{Helical Trilayer Graphene}\label{sec:intro_theory}

We first establish the topology of an individual HTG domain at the non-interacting level, for $\theta=\SI{1.79}{\degree}$~\cite{popov_magic_2023, devakul_magic-angle_2023, yang_multi-moire_2024}. 
We apply a generalized Bistritzer-MacDonald continuum model~\cite{bistritzer_moire_2011, devakul_magic-angle_2023} to calculate the local band structure on the moir\'e scale $\lambda_m\sim a/\theta\sim\SI{8}{nm}$ (here, $a=\SI{0.246}{nm}$ is the lattice constant of graphene); see Sec.~\ref{sup_sec:BM}.
Lattice relaxation favors an intermoir\'e shift that locally breaks $C_{2z}$ symmetry, forming two types of domains, h-HTG and $\bar{\text{h}}$-HTG (Fig.~\ref{fig:schematic}B), that are related to one another by a $C_{2z}$ transformation.
The two domain types tile the sample such that $C_{2z}$ symmetry is globally preserved. 
Within each domain, local breaking of $C_{2z}$ symmetry allows the moir\'e bands to carry valley-contrasting Chern numbers (Fig.~\ref{fig:schematic}C).
Because the domains are related by $C_{2z}$, the local bands in neighboring domains carry opposite Chern numbers within each flavor when time-reversal symmetry is unbroken, producing a topological mismatch across every domain wall.

Although an isolated domain is gapped at filling factors $\nu=-4$, $0$, and $4$ (carriers per moir\'e unit cell), this topological mismatch prevents the mosaic from being globally insulating at $B=0$.
Allowing for symmetry-breaking and flavor polarization at other integer fillings generally does not remove this mismatch (see Supplemental Sec.~\ref{sup_sec:chern}).
A domain-wall network exhibits non-universal transport that depends on the domain structure, contact geometry, and mode hybridization. 
Obtaining both vanishing longitudinal resistance and a quantized Hall resistance would therefore require fine tuning~\cite{Sayak}.

% ---------------------------------------------------------------------------
\begin{figure*}[t]
\centering
\includegraphics{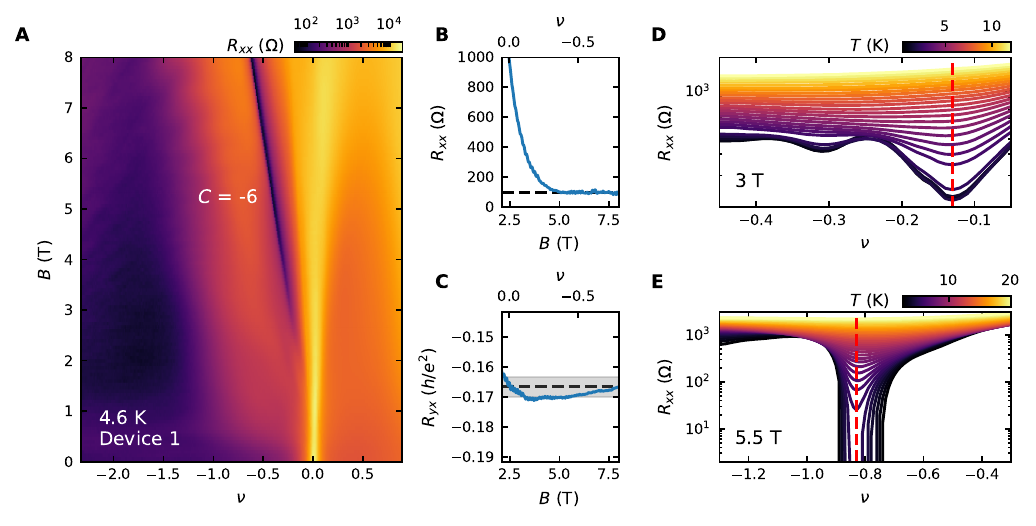}
\caption{\textbf{Magnetotransport of HTG.}
(\textbf{A}) Longitudinal resistance $R_{xx}$ as a function of $\nu$ and $B$ at $D=0$ and $T=\SI{4.6}{K}$. 
The sharp suppression of $R_{xx}$ emanating from the CNP exhibits a slope consistent with a $C=-6$ state.
(\textbf{B,C}) $R_{xx}$ (\textbf{B}) and $R_{yx}$ (\textbf{C}) interpolated along the dip in $R_{xx}$ of (\textbf{A}). 
The horizontal dashed lines indicate $\SI{100}{\Omega}$ and $-h/6e^2$ in (\textbf{B}) and (\textbf{C}), respectively. The gray shaded region in (\textbf{C}) indicates a $2$\% window around $-h/6e^2$. 
(\textbf{D}) Temperature dependence of $R_{xx}$ from $\SI{0.5}{K}$ to $\SI{12}{K}$ in $\sim\SI{0.5}{K}$ steps as a function of $\nu$ at $B=\SI{3}{T}$.
(\textbf{E}) Same as (\textbf{D}) for $T$ from $\SI{2.6}{K}$ to $\SI{20}{K}$ and $B=\SI{5.5}{T}$. 
In (\textbf{D}) and (\textbf{E}), vertical dashed red lines indicate $\nu$ where $C=-6$ is expected for the given magnetic field. 
Data in (\textbf{D}) and (\textbf{E}) are taken from a cool-down separate from that of (\textbf{A}). 
See Fig.~\ref{sup_fig:arrhenius} for fits to Arrhenius activation. 
}
\label{fig:high_temp_transport}
\end{figure*}
% ---------------------------------------------------------------------------

Real HTG devices need not form the ideal periodic domain pattern of Fig.~\ref{fig:schematic}B. 
Heterostrain---relative strain between layers---is typically present on the order of $0.1\%$~\cite{wangUnusualMagnetotransportTwisted2023, kazmierczakStrainFieldsTwisted2021, kerelsky2019maximized}.
Because the supermoir\'e is a higher-order modulation, it is exceptionally sensitive to heterostrain; imaging has shown that the structure can be substantially distorted and even changed between successive cool-downs~\cite{hoke_imaging_2024}.
Using torsional force microscopy (TFM)~\cite{pendharkar_torsional_2024}, we image a strongly distorted supermoir\'e in a separate HTG sample from the one used for transport measurements (Fig.~\ref{fig:schematic}D).
The domain walls are most clearly visible as variations in the extracted local moir\'e period (Fig.~\ref{fig:schematic}E; see Sec.~\ref{sup_sec:TFM}).
While such strain distorts the structure of the supermoir\'e, it is not expected to alter the Chern numbers of individual domains, leaving the number of modes at domain walls unchanged.
If each domain is large compared to the lateral confinement of the edge states, counter-propagating modes within a given flavor sector at opposing boundaries of a domain remain spatially separated and cannot annihilate.

%%%%%%%%%%%%%%%%%%%%%%%%%%%%%%%%%%%%%%%%%%%%%%%%%%%%%%%%%
\section*{Magnetotransport in HTG}\label{sec:main_transport}

The HTG device that is the primary focus of this work, previously studied in Ref.~\cite{xia_topological_2025}, has an interlayer twist of $\theta=1.79^\circ$, extracted from a full-filling density of $n_{\nu=\pm 4}=\pm7.45\, \times\, 10^{12}$ \si{\per\centi\metre\squared}.
At moir\'e fillings $\nu=2/3$, $1$, and $3$, correlated states exhibited an AHE with anomalous Hall resistances $|R_{yx}^{\mathrm{AH}}| \lesssim \SI{2}{k\Omega}$ concurrently with longitudinal resistances $R_{xx}\gtrsim \SI{2}{k\Omega}$~\cite{xia_topological_2025}, consistent with transport through a domain-wall network.

In a Landau fan measured at displacement field $D=0$ and $T=\SI{4.6}{K}$, we observe a sharp suppression of $R_{xx}$ that emanates from the CNP for fields above $\SI{2.5}{T}$ (Fig.~\ref{fig:high_temp_transport}A).
The trajectory of this resistance minimum follows a St\v{r}eda slope of $(h/e)(\partial n/\partial B)=-5.99\pm 0.02$, identifying a $C=-6$ state.
Along this trajectory, $R_{xx}$ falls to $\sim\SI{100}{\Omega}$ above $\SI{5}{T}$ and $R_{yx}$ approaches $-h/6e^2$ to within $2\%$ (Figs.~\ref{fig:high_temp_transport}B, C).
Quantization does not improve upon further cooling: at $\SI{300}{mK}$ and below, residual deviations persist, likely dominated by bulk conduction that mixes longitudinal and Hall signals (see Sec.~\ref{sup_sec:sym}).
The state is thermally activated, with Arrhenius gaps that increase from $1.17\pm 0.04$\,meV at $\SI{3}{T}$ to $4.86\pm 0.05$\,meV at $\SI{5.5}{T}$ (Figs.~\ref{fig:high_temp_transport}D, E and Fig.~\ref{sup_fig:arrhenius}).

\begin{figure*}[t]
\centering
\includegraphics{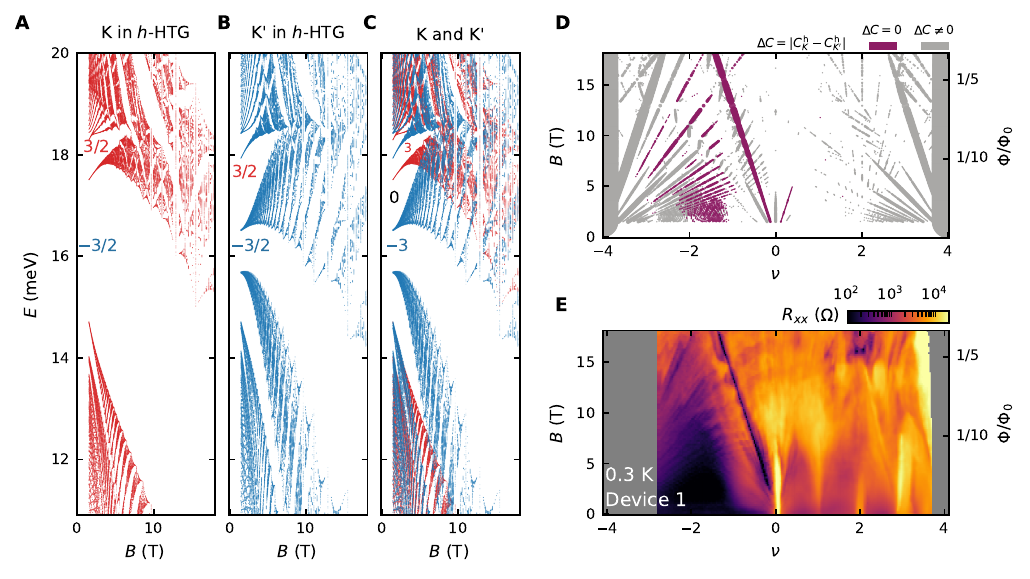}
\caption{\textbf{Hofstadter Spectrum of HTG.} 
(\textbf{A}) Calculated Hofstadter spectrum for valley $K$ in $\SI{1.79}{\degree}$ h-HTG (equivalently valley $K'$ in $\bar{\text{h}}$-HTG). 
The Chern numbers of selected gaps that emanate from the CNP are indicated, not accounting for spin degeneracy.
(\textbf{B}) Same as (\textbf{A}) for valley $K'$ in h-HTG (equivalently valley $K$ in $\bar{\text{h}}$-HTG).
(\textbf{C}) The combined spectra of both valleys in h-HTG.
(\textbf{D}) Wannier diagram corresponding to (\textbf{C}). The size of each dot corresponds to the size of the gap in the spectrum.
Each gap is color-coded according to Eq.~\ref{eq:global_gap_condition}, evaluated within h-HTG as the difference between its two valleys, $\Delta C = \left|C^{\text{h}}_{K} - C^{\text{h}}_{K'}\right|$.
Magenta dots indicate gaps where h-HTG and $\bar{\text{h}}$-HTG have the same flavor-resolved Chern numbers, and are therefore candidates for global gaps. 
Gray dots correspond to cases where h-HTG and $\bar{\text{h}}$-HTG have differing flavor-resolved Chern numbers and therefore a network of edge modes is expected.
(\textbf{E}) Longitudinal resistance $R_{xx}$ as a function of $\nu$ and $B$ at $D=0$ and $T=\SI{300}{mK}$.}
\label{fig:hofstadter}
\end{figure*}

Features emanating from $\nu=-4$ that are faint at \SI{4.6}{K} become more clearly resolved at \SI{300}{mK} (Fig.~\ref{fig:hofstadter}E), but remain poorly quantized (Fig.~\ref{sup_fig:minus4_fan}). 
This remains the case at lower temperatures, $T=\SI{37}{mK}$ (Fig.~\ref{sup_fig:dilfan}).
Line cuts of $R_{xx}$ and $R_{yx}$ at $\SI{8}{T}$ across these temperatures show that the $C=-6$ state is the only feature to exhibit concurrent low $R_{xx}$ and nearly quantized $R_{yx}$ (Fig.~\ref{sup_fig:alltemps_cut}).
We return to the features emanating from $\nu=-4$ below, where we argue that their poor quantization likely reflects twist-angle disorder.
In Sec.~\ref{sup_sec:dfield} we discuss the behavior at finite displacement field, where we observe $C=-6$, $-10$, $-14$ states emanating from the CNP.

% ---------------------------------------------------------------------------
% ---------------------------------------------------------------------------

Although a network of edge modes can yield quantized $R_{yx}$ or vanishing $R_{xx}$ in fine-tuned circumstances~\cite{Sayak}, similar $C=-6$ features in other devices (see Sec.~\ref{sup_sec:other_devices}) and the absence of other well-quantized features in transport suggest that this explanation is unlikely.
Instead, we propose that both $\text{h}$-HTG and $\bar{\text{h}}$-HTG develop a global $C=-6$ gap with consistent Chern numbers within each flavor across domains, eliminating the need for interior edge modes. 
We now demonstrate that this can occur near the CNP in the presence of a magnetic field using two complementary calculations: the full Hofstadter spectrum of HTG and a simplified orbital Zeeman model that isolates the mechanism of the transition.

%%%%%%%%%%%%%%%%%%%%%%%%%%%%%%%%%%%%%%%%%%%%%%%%%%%%%%%%%
\section*{Hofstadter Spectrum}\label{sec:main_hofstadter}
 
We begin by computing the Hofstadter spectra for both valleys of a $\theta=1.79^\circ$ h-HTG domain. 
Because $C_{2z}$ is broken within each domain, the two valleys are inequivalent and must be treated separately. 
We label each Hofstadter gap by $C_{\tau,s}$, the total Chern number of the occupied bands for valley $\tau$ and spin $s$.
At low fields, the dominant gaps that emanate from the CNP have $C_{\tau, s}=\pm3/2$ per spin and valley (Figs.~\ref{fig:hofstadter}A,B), reflecting the chirality of the three Dirac cones gapped within each flavor, each contributing $\pm 1/2$.
Therefore, the integer Chern number of the central valence band combines with a formally half-integer contribution from the remote valence bands~\cite{nakatsuji2023multiscale, kwan2024strong, haldane_model_1988}.
Crucially, in the limit of conserved valley, a globally gapped state requires  the two domains to have identical Chern numbers per flavor: 
\begin{equation}\label{eq:global_gap_condition}
\Delta C = \left|C^{\text{h}}_{K, s} - C^{\bar{\text{h}}}_{K, s}\right| =
\begin{cases}
0  & \text{global gap possible},\\
>0  & \text{domain-wall modes}.
\end{cases}
\end{equation}
Because $C_{2z}$ maps one domain onto the other while exchanging the valleys, $C^{\bar{\text{h}}}_{K, s} = C^{\text{h}}_{K', s}$; $\Delta C$ can therefore be evaluated within a single domain by comparing the Chern numbers of its two valleys (Figs.~\ref{fig:hofstadter}A,B).

% ---------------------------------------------------------------------------
\begin{figure*}[t]
\centering
\includegraphics{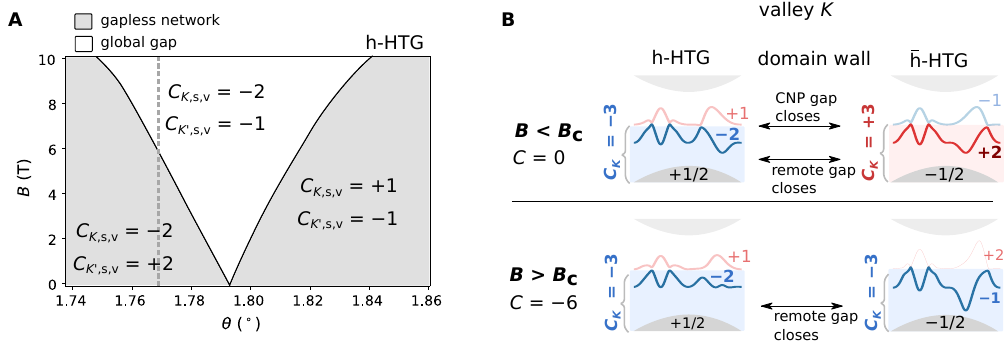}
\caption{\textbf{Orbital Zeeman model.}
(\textbf{A}) Chern number of the central valence band in each flavor as a function of $B$ and $\theta$ for h-HTG, within the orbital Zeeman model. 
Gray (white) region corresponds to a total Chern number $C=0$ ($C=-6$) of the central valence bands including spin, and leads to a gapless network (globally gapped state).
(\textbf{B}) Schematic comparison of Chern numbers in valley $K$ between h-HTG and $\bar{\text{h}}$-HTG domains.
The central bands are labeled with their individual Chern numbers.
We also indicate the half-integer values for the set of remote valence bands. 
$C_{2z}$ fixes the Chern numbers of valley $\tau$ in h-HTG to be identical to those of valley $-{\tau}$ in $\bar{\text{h}}$-HTG.
At $B=0$, the total valley-resolved Chern number at the CNP varies between h and $\bar{\text{h}}$ domains, forming gapless modes on the domain wall at the CNP.
At $B=B_c$, a topological transition within the domains occurs only in valley $K$ of $\bar{\text{h}}$-HTG and valley $K'$ of h-HTG, leading to $C=-6$ throughout the sample.
For $B>B_c$, the total valley-resolved Chern numbers of the CNP gap, which include a factor of two from spin, are identical for both domains, removing the topological protection of the edge modes and enabling a gapped domain wall at the CNP.
Note that a closure of the remote valence gap at the domain wall is still required. 
Dispersions of the central bands are computed using the orbital Zeeman model at $\theta=1.77^\circ$ (see dashed line in \textbf{A}) for $B=\SI{0}{T}$ ($B<B_c$) and $\SI{11}{T}$ ($B>B_c$).
}
\label{fig:gfactor}
\end{figure*}
% ---------------------------------------------------------------------------

Absent flavor polarization, a domain is gapped only when both valleys are simultaneously gapped.
Figure~\ref{fig:hofstadter}C shows the superposition of the two spectra.
The $C_{\tau, s}=\pm3/2$ Hofstadter gaps of the two valleys combine to form gaps with total Chern number per spin $C_s=0$, $3$, and $-3$ ($C=0$, $6$, and $-6$ including spin degeneracy). 
The $C_s=0$ gap is not a global gap: it hosts a shunting network of gapless modes because $C^{\text{h}}_{K, s}= -C^{\bar{\text{h}}}_{K, s}=-3/2$.
The $C_s=\pm3$ gaps satisfy the matching condition of Eq.~\ref{eq:global_gap_condition} across the two domain types, $\Delta C=0$, so no protected domain-wall modes are required.
While the $C_s=3$ gap closes near $\SI{4.5}{T}$, the $C_s=-3$ gap grows with field, in agreement with our experimental observations.
The calculated $C_s=-3$ gap sizes are $0.8$ and $\SI{1.4}{meV}$ at $\SI{3}{T}$ and $\SI{5.5}{T}$, respectively. 
These values are consistently smaller and grow more slowly with field than what we observe experimentally (Fig.~\ref{sup_fig:arrhenius}).
This may indicate that the microscopic parameters of the model are inaccurate, or that interactions further stabilize the gap~\cite{nuckolls2025spectroscopy, wang_theory_2024}.

To compare the calculated gaps directly with the experimental Landau fan, we represent them in a Wannier diagram~\cite{wannierResultNotDependent1978, hejazi_landau_2019}, fully accounting for all spins and valleys (Fig.~\ref{fig:hofstadter}D). 
Each gap appears as a dot whose size reflects the gap magnitude and whose color indicates whether it satisfies the global-gap condition of Eq.~\ref{eq:global_gap_condition}.
We find a prominent global $C=-6$ Hofstadter gap emanating from the CNP, in agreement with experiment. 
Our calculations further predict that this gap persists for a range of twist angles $\SI{1.7}{\degree} \lesssim \theta \lesssim \SI{1.9}{\degree}$ (Sec.~\ref{sup_sec:hof}).

While the existence of the global gap is robust, the finer features of the calculated spectra are sensitive to the twist angle. 
At $\theta=\SI{1.79}{\degree}$ the calculated $C=-6$ gap extends below \SI{1}{T}, while at $\theta=\SI{1.75}{\degree}$ it onsets near \SI{5}{T} (Fig.~\ref{sup_fig:1p75_hoff}), bracketing the experimental onset of $B\approx\SI{2.5}{T}$.
Similarly, the small $C=+6$ gap present in the calculated $\theta=\SI{1.79}{\degree}$ spectra, which we do not observe in experiment, is absent at $\theta=\SI{1.75}{\degree}$.
Slightly tuning the microscopic parameters of the model at $\theta=\SI{1.79}{\degree}$ can produce similar effects.
We also note that for electron doping, this sample exhibits isospin-symmetry-breaking order~\cite{xia_topological_2025,bocarsly2026reversals} such that our single-particle calculations are no longer fully applicable.

Figure~\ref{fig:hofstadter}D reveals a series of candidate global gaps with $C=10,$ $14,$ $18$,... emanating from $\nu=-4$. 
While we observe corresponding suppressions in $R_{xx}$ (Fig.~\ref{fig:hofstadter}E), they neither follow integer-sloped St\v{r}eda lines nor show well-quantized transport, likely due to the sensitivity of states emanating from nonzero filling to twist-angle disorder~\cite{Uri2020mapping} (see Sec.~\ref{sup_sec:sym}).

Under the application of a displacement field, we observe experimentally that the $C=-6$ state is joined by other states with $C=-10,-14$, in agreement with our Hofstadter calculations (Sec.~\ref{sup_sec:dfield}). 
Each state corresponds to emptying one additional Landau level per flavor, decreasing the total Chern number by four while preserving $\Delta C=0$.

%%%%%%%%%%%%%%%%%%%%%%%%%%%%%%%%%%%%%%%%%%%%%%%%%%%%%%%%%
\section*{Orbital Zeeman model}\label{sec:main_gfactor}

To isolate the mechanism of the field-induced valley-selective transition, we consider a simplified model in which the magnetic field enters solely via the orbital Zeeman effect~\cite{Slizovskiy2019Films, Hwang2021Geometric, Auerbach2025Isospin}.
While this approach neglects Hofstadter physics and is therefore applicable only to low fields, it preserves translation symmetry, allowing for a Bloch-bands description.
The strength of the coupling is set by the valley-contrasting $g$-factor $g_{\tau}(\bm{k})$~\cite{sun2020topologicalzeeman, zhang_spin-polarized_2022}, which is a matrix we compute in the space of the central bands---the two bands closest to the CNP---of each domain (Sec.~\ref{sup_sec:gk_calcs}).
Because this coupling is momentum dependent, an applied field distorts the bands, redistributing Berry curvature within each band.
Furthermore, the matrix structure of $g_{\tau}(\bm{k})$  can mix the central bands: a sufficiently strong field closes and reopens the CNP gap, allowing Berry curvature to transfer between the bands.

By solving the continuum model with the additional term $-g_{\tau}(\bm{k})\mu_B B$, we extract the Chern number $C_{\tau,s,\text{v}}$ of the central valence (v) band within one flavor, as shown in Fig.~\ref{fig:gfactor}A for h-HTG. 
At zero field, the central bands undergo a topological transition near the magic angle, changing from $C_{K,s,\text{v}}=-2$ to $+1$ as $\theta$ is increased through $\sim\SI{1.8}{\degree}$; time-reversal symmetry fixes the other valley to $C_{K',s,\text{v}}=-C_{K,s,\text{v}}$.
The experimental device therefore lies close to a topological phase boundary. 
These Chern numbers are identical to those of the sublattice-polarized bands which enter the strong-coupling theory of HTG~\cite{devakul_magic-angle_2023, kwan2024strong, xia_topological_2025}. 
A magnetic field shifts the critical angle for the transition in opposite directions between the two valleys; at a fixed twist angle, one valley can undergo the transition as a function of field while the other does not.
Therefore, our calculation finds a window of magnetic fields and twist angles where $C_{K,s,\text{v}}+C_{K',s,\text{v}}=-3$.
Because the remote valence gap has a total Chern number of zero when summed over all flavors, this is consistent with the observed $C=-6$ state once spin degeneracy is included. 

We now revisit the global-gap condition for the CNP gap.
The relevant flavor-resolved Chern numbers $C_{\tau, s} = \pm 3/2$ of the CNP gaps differ from those of the central bands because of the contribution of the remote valence bands.
Figure~\ref{fig:gfactor}B illustrates the two regimes.
Below the transition field $B_c$, a given valley has opposite valley-resolved Chern numbers in the two domain types ($\Delta C\ne 0$), forcing gapless helical valley modes at every domain wall.
Above $B_c$, the total Chern number of the occupied bands within each valley is identical in the two domain types ($\Delta C = 0$), and the domain walls need not host gapless modes.
This does not mean that every band has become topologically identical: the central valence band still carries different Chern numbers in the two domain types.
In the $C=-6$ state, the central valence band of valley $K$ carries Chern number $-2$ in h-HTG and $-1$ in $\bar{\text{h}}$-HTG (with the values interchanged for valley $K'$).
This mismatch in Chern number of the central valence band in a given flavor across a domain wall is absorbed by the remote valence bands, the gap to which closes at the domain walls.
Therefore, the topology can rearrange away from the Fermi level while the CNP gap remains open everywhere.

%%%%%%%%%%%%%%%%%%%%%%%%%%%%%%%%%%%%%%%%%%%%%%%%%%%%%%%%%
\section*{Discussion}\label{sec:main_discussion}

Our results demonstrate how a Chern mosaic can become a global Chern insulator without eliminating its underlying spatially varying band topology.
In HTG, the zero-field Chern mosaic is enforced by $C_{2z}\mathcal{T}$, which relates the two domain types within a given valley with an accompanying reversal of their band Chern numbers.
This generally produces a Chern-number mismatch across domain walls, and hence gapless modes.
Indeed, the anomalous Hall states in HTG are consistent with a domain-wall network~\cite{xia_topological_2025}.
A magnetic field lifts this constraint: the two domain types remain related only by $C_{2z}$, which exchanges valleys, but their valley Chern numbers are no longer constrained to be opposite.
The Chern number below a given gap can therefore become identical across domains even while the individual bands retain different Chern numbers. 
The $C=-6$ state emanating from charge neutrality realizes this possibility.
The Chern-number mismatch of the central bands is compensated by the remote valence bands, the gap to which closes at the domain walls while allowing the CNP gap to remain open.
Thus, a global Chern gap can emerge without erasing the underlying band-resolved Chern mosaic.

Our calculations place the device close to a topological phase boundary, naturally explaining why a moderate magnetic field suffices to drive the transition. 
The magnetic field is not unique in this respect: an interaction-driven state that both breaks time-reversal symmetry and drives the same valley-selective topological transition could produce a global Chern insulator at zero applied field, leading to a quantum anomalous Hall effect~\cite{Sharpe2019, Serlin2020-sr}.
A related mechanism has been proposed in rhombohedral pentalayer graphene, where a displacement field drives a band inversion in a spin-valley flavor of a correlation-driven insulating state, producing a $C=-5$ state that is stabilized experimentally by a small magnetic field~\cite{han_correlated_2024}.
In HTG, by contrast, the transition must not only invert bands within each domain but also eliminate the flavor-resolved Chern mismatch between the two domain types.

The impact of domains on transport in Chern insulators is dictated not by their mere presence, but by how the topological invariants change across domain boundaries.
\ce{Bi2Te3}-class topological insulators, for example, contain abundant crystallographic twin domains~\cite{kriegner_twin_2017} while exhibiting well-quantized transport~\cite{fox_part-per-million_2018, he2013review}, because the symmetry operation relating the twin domains preserves the Chern number of the magnetically gapped surface state~\cite{andersen_topological_2023}.
Chern mosaics also arise in magic-angle twisted bilayer graphene coupled to hBN, where spatial variations in the graphene-hBN registry can change the local Chern number~\cite{grover_chern_2022}. 
Unlike in HTG, the neighboring regions are not a pair of symmetry-related structural domains with symmetry-enforced Chern numbers.
Whether such a mosaic can be tuned into a global Chern state remains open: it would require a tuning parameter that eliminates the Chern-number mismatch below the relevant gap across the sample.
Other supermoir\'e platforms---including unequal-twist trilayer graphene~\cite{uri2023quasicrystalSC, mao2023supermoireffective, nakatsuji2023multiscale, foo_extended_2024, yang_multi-moire_2024, hao2024robust, xia2025magiccontinuummultimoiretwisted, hoke_linking_2025,Zhou2026supermoire}, helical quadrilayer graphene~\cite{fujimoto2025quadrilayer}, and trilayer transition metal dichalcogenides~\cite{nakatsuji2025moirebandengineeringtwisted, choi2025higherchernbandshelical, beach2026electricallyprogrammable, wang2026supermoirechernmosaic}---may hinge on a similar interplay between the relaxed structure and its local symmetries.

In HTG, the degeneracy of the network may change at positive commensurate fillings, where correlated states have previously been observed~\cite{xia_topological_2025}.
At fractional moir\'e fillings, theory predicts a network of fractional Chern-insulator edge states~\cite{kwan2024fractionalchernmosaicsupermoire}.
Such edge-mode networks can exhibit a variety of contact-dependent resistances~\cite{Sayak}, and their Luttinger-liquid properties, edge equilibration, and junction scattering could be probed using interior contacts~\cite{wang2022tWTe2} or scanning-probe microscopy. 
The geometry of the domain-wall network is set by the supermoir\'e structure and is therefore highly sensitive to strain, which provides a natural means to reshape it (see Sec.~\ref{sup_sec:twist}).
Together with magnetic field and doping, this tunability makes HTG a platform for controlled studies of chiral and helical edge-mode physics.

%%%%%%%%%%%%%%%%%%%%%%%%%%%%%%%%%%%%%%%%%%%%%%%%%%%%%%%%%
\textbf{Acknowledgments}
We thank Sayak Bhattacharjee, Jesse Hoke, Yifan Li, Yuwen Hu, Patrick Ledwith, Sandesh Kalantre, Ben Feldman, Sergio de la Barrera, Dan Parker, Mikito Koshino, Oskar Vafek, Marc Kastner, and David Goldhaber-Gordon for fruitful discussions. 
We thank Rupini Kamat for assistance in dilution refrigerator measurements.
We thank A. Bangura, G. Jones, R. Nowell, A. Woods, and S. Hannahs for supporting measurements performed at the National High Magnetic Field Laboratory.
Work in the P.J.-H. group was partially supported by the Army Research Office MURI W911NF2120147, the Air Force Office of Scientific Research (AFOSR) grant FA9550-21-1-0319, the National Science Foundation (DMR-1809802), a Max Planck-Humboldt Research Award to P.J.-H., the CIFAR Quantum Materials Program, the Ramón Areces Foundation, and the Gordon and Betty Moore Foundation’s EPiQS Initiative through grant GBMF9463 to P.J.-H.
L.-Q.X. acknowledges support from the MathWorks Fellowship.
A.U. acknowledges support from the MIT Pappalardo Fellowship and from the VATAT Outstanding Postdoctoral Fellowship in Quantum Science and Technology.
Analysis and measurements performed by A.S. were supported by the US Department of Energy, Office of Science, Basic Energy Sciences, Materials Sciences and Engineering Division, under Contract DE-AC02-76SF00515.
K.W. and T.T. acknowledge support from the JSPS KAKENHI (Grant Numbers 21H05233 and 23H02052) , the CREST (JPMJCR24A5), JST and World Premier International Research Center Initiative (WPI), MEXT, Japan.
T.D. acknowledges support from the Air Force Office of Scientific Research
under award number FA9550-25-1-0343.
A portion of this work was performed at the National High Magnetic Field Laboratory, which is supported by National Science Foundation Cooperative Agreement No. DMR-2128556 and the State of Florida.
This work was performed in part at the Harvard University Center for Nanoscale Systems (CNS); a member of the National Nanotechnology Coordinated Infrastructure Network (NNCI), which is supported by the National Science Foundation under NSF award no. ECCS-2025158.
This work was carried out in part through the use of MIT.nano's facilities.

%%%%%%%%%%%%%%%%%%%%%%%%%%%%%%%%%%%%%%%%%%%%%%%%%%%%%%%%%
\textbf{Author Contributions}

L.-Q.X., A.U., and A.S. conceived the project.
L.-Q.X. fabricated the devices with the help of A.U. 
L.-Q.X. and A.S. carried out the TFM measurements.
L.-Q.X. and A.S. carried out the dilution refrigerator measurements. 
L.-Q.X. and A.U. carried out the measurements at the National High Magnetic Field Laboratory. 
L.-Q.X. and A.S. carried out helium-3 measurements.
J.M.-M., T.D., and Y.H.K. performed the Hofstadter and orbital Zeeman calculations.
K.W. and T.T. supplied the boron nitride crystals.
A.S. analyzed the data with the help of L.-Q.X., A.U., Y.H.K., T.D., Z.W.G., and M.P.A., and input from all authors.
A.S. and Y.H.K. wrote the manuscript with input from all authors.
P.J.-H. and A.S. supervised the project.

\section*{Competing interests}

The authors declare no competing interests.

\section*{Data availability}

The experimental data supporting the findings of this study are available at the Stanford Digital Repository~\cite{data_repo}. The code used to generate the figures from these data is available at \url{https://github.com/sharpelab/Quantized-Transport-through-a-Supermoire}. Code for the Hofstadter and orbital Zeeman calculations is available from the corresponding authors upon reasonable request.

%%%%%%%%%%%%%%%%%%%%%%%%%%%%%%%%%%%%%%%%%%%%%%%%%%%%%%%%%
\onecolumngrid
\bibliography{bib}

%%%%%%%%%%%%%%%%%%%%%%%%%%%%%%%%%%%%%%%%%%%%%%%%%%%%%%%%%
\clearpage
\renewcommand\thefigure{S\arabic{figure}}
\setcounter{figure}{0}
\renewcommand\thetable{S\arabic{table}}
\setcounter{table}{0}
\renewcommand\thesubsection{S\arabic{subsection}}
\setcounter{subsection}{0}
\section*{Supplemental material}

%%%%%%%%%%%%%%%%%%%%%%%%%%%%%%%%%%%%%%%%%%%%%%%%%%%%%%%%%
\subsection{Methods} \label{sup_sec:fab}

Van der Waals heterostructures were assembled using standard dry-transfer techniques, following the procedures described in Refs.~\cite{xia_topological_2025,Sun2025Optimized}.

Torsional force microscopy was performed on a Bruker Icon XR atomic force microscope using Adama Innovations AD-2.8-SS tips. 
The tips were mounted in a DTRCH-AM torsional probe holder to enable excitation of torsional modes. 
The torsional resonance used for imaging occurs near $\SI{1.25}{MHz}$. 
Further details of the imaging procedure can be found in the supplemental materials of Ref.~\cite{pendharkar_torsional_2024}.

Low-temperature electrical transport measurements were carried out in a Janis helium-3 refrigerator, the SCM-1 dilution refrigerator at the National High Magnetic Field Laboratory, and a Leiden Cryogenics CF-900 dilution refrigerator.
Details can be found in Ref.~\cite{xia_topological_2025}.

The dual gated geometry of the device allows for independent control of the electron density, $n$, and the perpendicular electric displacement field, $D$. Considering a parallel plate capacitor model, $n=(\epsilon_\mathrm{BN}\epsilon_0/e$)($V_\mathrm{bg}$/$d_\mathrm{bg} + V_\mathrm{tg}$/$d_\mathrm{tg}$) and $D=(\epsilon_\mathrm{BN}\epsilon_0/2$)($V_\mathrm{bg}$/$d_\mathrm{bg} - V_\mathrm{tg}$/$d_\mathrm{tg}$).
Here, $V_{bg}$ ($V_{tg}$) is the voltage applied to the bottom (top) gate, $\epsilon_\mathrm{BN}=3$ is the relative dielectric constant of hBN, $\epsilon_0$ is the vacuum permittivity, $e$ is the elementary charge, and $d_\mathrm{bg}$ ($d_\mathrm{tg}$) is the thickness of the bottom (top) hBN.

%%%%%%%%%%%%%%%%%%%%%%%%%%%%%%%%%%%%%%%%%%%%%%%%%%%%%%%%%
\subsection{Twist Angle Extraction and the Impact of Strain} \label{sup_sec:twist}

The extraction of twist angles for the main device was previously discussed in Ref.~\cite{xia_topological_2025}.
We repeat in brief the salient points here and expand upon the ramifications of strain in this extraction.
Continuum model calculations of h-HTG and $\bar{\text{h}}$-HTG find large band gaps at $\nu=\pm4$, corresponding to four electrons (holes) per moir\'e unit cell.
At $\nu=\pm4$, as discussed in the main text, we expect the domain walls to form a network of edge modes
that shunts the gapped domains, lowering the resistance at these fillings compared to a homogeneous insulator.
Nevertheless, we are able to identify resistive peaks at $\nu=\pm4$ and signatures of Landau levels emerging from the band extrema.
Ignoring any contributions from quantum capacitance, we then fit a series of integer slopes to the measured Landau level gaps (dips in $R_{xx}$ emanating from $\nu=-4, 0, +4$) and resistive states at partial fillings ($R_{xx}$ peaks at $\nu=1,2,3$), using the density $n_s$ as a free parameter.
The best fit across all fillings and sloped features yields $4n_s = \SI{7.45 \pm 0.17e12}{cm^{-2}}$. Errors for this extraction are estimated by aligning the collection of $R_{xx}$ features to the left and right edges of each feature ($R_{xx}$ minima for Landau level gaps or peaks of correlated states).

The quantity $n_s$ is a measure of the carrier density that corresponds to a filling of one electron per moir\'e unit cell and does not yield any information about the shape of the moir\'e unit cell.
Despite the likely presence of strain within the sample, it is standard practice to assume zero strain so that $4n_s = \pm 32\sin^{2}(\theta/2)/\sqrt{3}a^2 \approx \pm 8\theta^2/\sqrt{3}a^2$, where $a=\SI{0.246}{nm}$ is the lattice constant for graphene and $\theta$ is the interlayer twist.
Under this assumption, the density at the extracted resistance peak therefore corresponds to $\theta = \SI{1.79 \pm 0.02}{\degree}$.  

During device fabrication, we target $\theta_{12}=\theta_{23}$.
In the likely case of a slight mismatch between the two twist angles, $\theta_{12} \neq \theta_{23}$, theoretically we expect that the system relaxes to a structure similar to that of the equiangle one, only with a smaller supermoir\'e unit cell~\cite{xia_topological_2025}.
Small variations in $\theta_{12}$ or $\theta_{23}$ will result in large variations in the supermoir\'e wavelength, but lattice relaxation calculations suggest (distorted) h-HTG and $\bar{\text{h}}$-HTG supermoir\'e domains will still form regardless of their size~\cite{xia_topological_2025, hoke_imaging_2024}.
Indeed, within our range of accessible gate voltages, we only observe signatures of a single moir\'e periodicity. 
We interpret the phenomena discussed in the main text as arising from the existence of moir\'e-periodic domains separated by structural domain walls. 
Therefore, even with deviations from exactly equal twist angles, while the supermoir\'e pattern may change significantly, the physical properties of HTG will remain robust so long as the details of the moir\'e-scale bands representative of the domains do not change significantly. 
We are unable to resolve any transport features which correspond to a supermoir\'e unit cell area, possibly due to substantial variation of the supermoir\'e throughout the device. 

The above discussion assumed zero strain despite the fact that moir\'e samples typically have strains of magnitude $0.1-0.7\%$~\cite{kerelsky2019maximized, choi2019electronic, xie2019spectroscopic, mesple_heterostrain_2021, finney_extended_2025}.
We will now calculate the expected correction to the extracted twist angle under the influence of strain.
Strain is described by a rank-2 tensor $S_{\epsilon}$, 
\begin{align}
    S_{\epsilon} & = \begin{pmatrix} -\epsilon_\mathrm{uni} + \epsilon_\mathrm{bi} & \epsilon_{\mathrm{shear}}\\
\epsilon_{\mathrm{shear}} & \nu\epsilon_\mathrm{uni} + \epsilon_\mathrm{bi} \end{pmatrix}  \  ,
\end{align}
where $\epsilon_{\mathrm{uni}}$, $\epsilon_{\mathrm{bi}}$, and $\epsilon_{\mathrm{shear}}$ correspond to the amount of uniaxial, biaxial, and shear strain, respectively. 
$\nu$ is the Poisson ratio. 
For graphene, $\nu\approx 0.16$~\cite{wangUnusualMagnetotransportTwisted2023}.
In general, each layer of a heterostructure can have a distinct strain tensor.
For homogeneous strain within a given layer, we can ignore the shear component $\epsilon_{\mathrm{shear}}$, because one can always rotate into the principal axes of strain where $\epsilon_{\mathrm{shear}}=0$. 

For a given strain amplitude, biaxial strain is energetically more costly because it produces pure areal dilatation and therefore couples solely to the large bulk modulus of the 2D material.
For the same amount of uniaxial strain, it decomposes into a smaller dilation component and a larger deviatoric component that couples to the significantly smaller shear modulus, yielding a lower elastic energy cost.

We begin by considering homostrain, where all layers are identically strained. If we assume the twist angle in the absence of strain is $\theta_0$, then the twist angle inferred from transport, $\theta_{\mathrm{obs}}$, in the presence of small biaxial homostrain $\epsilon_{\mathrm{bi}}$ is
\begin{equation}
    \theta_{\mathrm{obs}} = \theta_0(1-\epsilon_{\mathrm{bi}}+\dots),
\end{equation}
where we have retained the leading-order correction.
Similarly, for small uniaxial homostrain $\epsilon_{\mathrm{uni}}$, we have
\begin{equation}
    \theta_{\mathrm{obs}} = \theta_0\left(1-\frac{1-\nu}{2}\epsilon_{\mathrm{uni}}+\dots\right).
\end{equation}

The shape of the moir\'e unit cell, band structure, and other associated properties are more sensitive to heterostrain than homostrain. 
However, heterostrain is more complicated in the presence of three layers. If we simply consider straining a single layer relative to the other two (a situation considered in Ref.~\cite{finney_extended_2025}), then the correction to $\theta_{\mathrm{obs}}$ is now second order in the strain magnitude but magnified by $1/\theta_0^2$. Under the presence of small biaxial heterostrain $\epsilon_{\mathrm{bi}}$, the leading-order correction is
\begin{equation}
    \theta_{\mathrm{obs}}= \theta_0\left(1+\frac{\epsilon_{\mathrm{bi}}^2}{2\theta_0^{2}}+\dots\right).
\end{equation}
Similarly, for small uniaxial heterostrain $\epsilon_{\mathrm{uni}}$, we have
\begin{equation}
    \theta_{\mathrm{obs}} = \theta_0\left(1-\frac{\nu\epsilon_{\mathrm{uni}}^2}{2\theta_0^2}+\dots\right).
\end{equation}
Despite being a second order correction, for a fixed amount of strain, biaxial heterostrain can be the largest correction given that it is amplified by $1/\theta_0^2$. 
For the device presented in the main text with an extracted twist angle $\theta = \SI{1.79 \pm 0.02}{\degree}$, this strain-induced correction to $\theta_0$ for realistic amounts of strain is likely an order of magnitude smaller than our reported error extracted from the transport measurement.

The supermoir\'e structure is even more sensitive to strain than the underlying moir\'e unit cells. Following the analysis of Ref.~\cite{hoke_imaging_2024}, we compute the reciprocal lattice vectors of each graphene layer under the influence of strain. 
Using these, we determine the supermoir\'e reciprocal lattice vectors, which we then invert to obtain the real-space supermoir\'e lattice vectors.
In the following, we will consider magic-angle HTG with $\theta_{12}=\theta_{23}=\SI{1.8}{\degree}$. Throughout, we will use $\varepsilon_i$ to denote a strain applied to layer $i$. The type of strain will be explicitly stated in the text or caption. 
In the case of uniaxial strain, $\phi_i$ denotes the direction of the strain within the corresponding layer, where $\phi_i=\SI{0}{\degree}$ corresponds to the zig-zag direction of that layer ($\phi_i=\SI{30}{\degree}$ corresponds to the armchair direction). Without loss of generality, we assume $\phi\le30$; any angle outside of this range is equivalent to an angle within this range under the symmetries of the lattice.

\begin{figure*}[h]  
\centering
\includegraphics{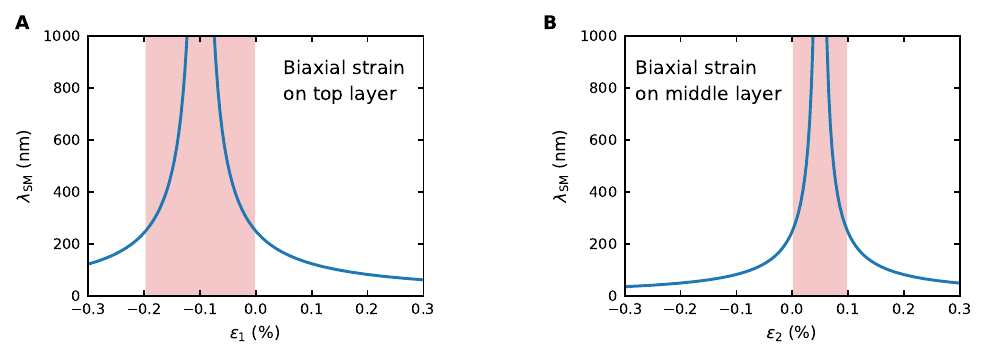}
\caption{\textbf{Effect of biaxial heterostrain on the supermoir\'e.} (\textbf{A}) Supermoir\'e wavelength $\lambda_{\mathrm{SM}}$ of $\SI{1.8}{\degree}$ HTG as a function of an isotropic biaxial heterostrain $\varepsilon_1$ applied to the topmost layer. This situation is equivalent to straining the bottommost layer. (\textbf{B}) Same as (\textbf{A}) but for an isotropic biaxial strain $\varepsilon_2$ applied to the middle layer. For both panels, the range of strains over which the supermoir\'e area is enhanced relative to the zero strain case is shaded in light red.}
\label{sup_fig:biaxial}
\end{figure*}

We will first consider an isotropic biaxial heterostrain applied to a single layer. As isotropic biaxial heterostrain does not break $C_{3z}$, the supermoir\'e wavevectors retain the same orientations and relative magnitudes. However, biaxial heterostrain can dramatically change the size of supermoir\'e domains (Fig.~\ref{sup_fig:biaxial}). For realistic amounts of strain, the supermoir\'e wavelengths can diverge, reflecting the fact that strain is driving the moir\'e patterns to be exactly commensurate. 

% ---------------------------------------------------------------------------
% ---------------------------------------------------------------------------

\begin{figure*}[h]
\centering
\includegraphics{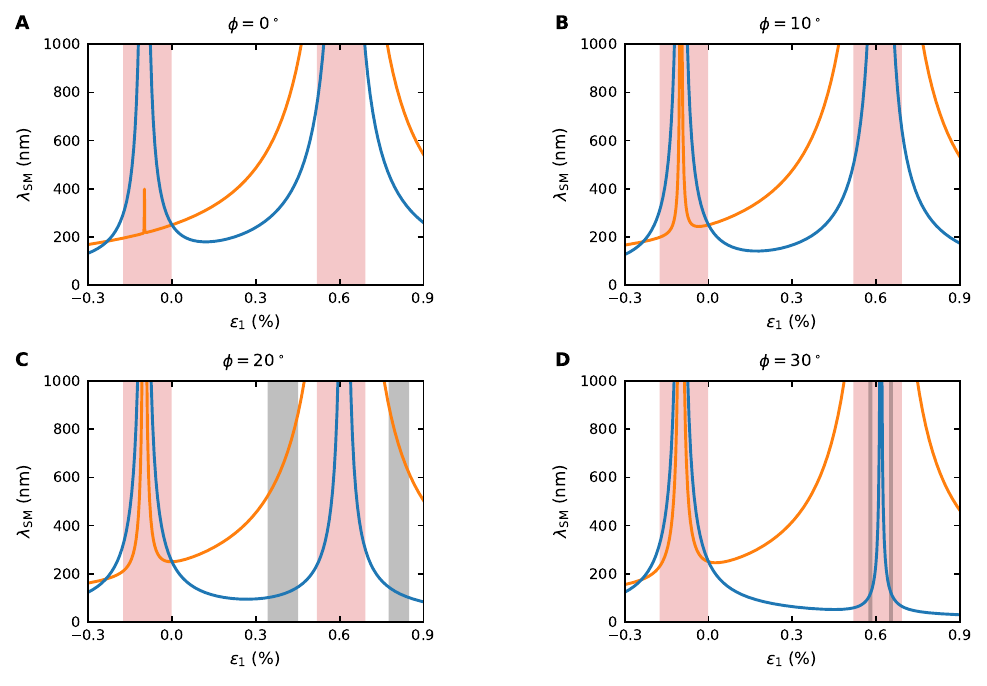}
\caption{\textbf{Effect of uniaxial heterostrain applied to the top layer on the supermoir\'e.} (\textbf{A}-\textbf{D}) Supermoir\'e wavelength $\lambda_{\mathrm{SM}}$ of $\SI{1.8}{\degree}$ HTG as a function of uniaxial heterostrain $\varepsilon_1$ applied to the topmost layer oriented along (\textbf{A}) $\phi_1=\SI{0}{\degree}$, (\textbf{B}) $\phi_1=\SI{10}{\degree}$, (\textbf{C}) $\phi_1=\SI{20}{\degree}$, and (\textbf{D}) $\phi_1=\SI{30}{\degree}$. The orange and blue curves correspond to the supermoir\'e wavelength along different directions. For all panels, the range of strains over which the supermoir\'e area is enhanced relative to the zero strain case is shaded in light red. We highlight in gray values of strain consistent with Fig.~\ref{fig:schematic}D and E, where the shorter supermoir\'e length satisfies $100\,\text{nm}\le\lambda_{\mathrm{SM}}\le150\,\text{nm}$ and the aspect ratio between the supermoir\'e lattice vectors $\ge4.5$.}
\label{sup_fig:uniaxial_top}
\end{figure*}

\begin{figure*}[h]
\centering
\includegraphics{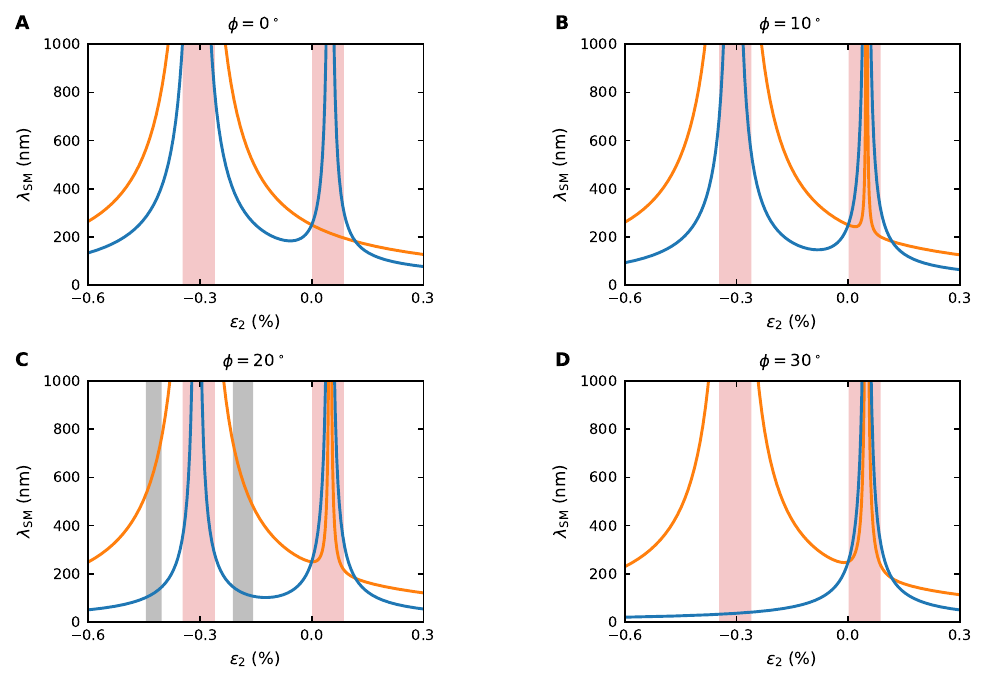}
\caption{\textbf{Effect of uniaxial heterostrain applied to the middle layer on the supermoir\'e.} (\textbf{A}-\textbf{D}) Supermoir\'e wavelength $\lambda_{\mathrm{SM}}$ of $\SI{1.8}{\degree}$ HTG as a function of uniaxial heterostrain $\varepsilon_2$ applied to the middle layer oriented along (\textbf{A}) $\phi_2=\SI{0}{\degree}$, (\textbf{B}) $\phi_2=\SI{10}{\degree}$, (\textbf{C}) $\phi_2=\SI{20}{\degree}$, and (\textbf{D}) $\phi_2=\SI{30}{\degree}$. The orange and blue curves correspond to the supermoir\'e wavelength along different directions. For all panels, the range of strains over which the supermoir\'e area is enhanced relative to the zero strain case is shaded in light red. We highlight in gray values of strain consistent with Fig.~\ref{fig:schematic}D and E, where the shorter supermoir\'e length satisfies $100\,\text{nm}\le\lambda_{\mathrm{SM}}\le150\,\text{nm}$ and the aspect ratio between the supermoir\'e lattice vectors $\ge4.5$.}
\label{sup_fig:uniaxial_mid}
\end{figure*}

Uniaxial heterostrain breaks $C_{3z}$, allowing for distortions of the supermoir\'e.
As with the biaxial case, the distortion to the supermoir\'e strongly depends on whether it is applied to the top layer (Fig.~\ref{sup_fig:uniaxial_top}) or middle layer (Fig.~\ref{sup_fig:uniaxial_mid}). 
Most notably, under uniaxial heterostrain the supermoir\'e wavelength can diverge along a single direction, leading to one-dimensional stripes in the supermoir\'e. In Fig.~\ref{fig:schematic}D and E, the supermoir\'e wavelength $\lambda_{\mathrm{SM}}\sim\SI{100}{nm}$ - $\SI{150}{nm}$ along the short axis with an aspect ratio $\gtrsim4.5$ (see Sec.~\ref{sup_sec:TFM} for comments regarding our ability to resolve features representative of the supermoir\'e). We highlight in gray the values of strain for which both of these criteria are satisfied. Though the parameter space is very large---each layer likely contains a slightly different amount of both uniaxial and biaxial heterostrain---we find a range of experimentally realistic strains consistent with the TFM image.

% ---------------------------------------------------------------------------
% ---------------------------------------------------------------------------

% ---------------------------------------------------------------------------
% ---------------------------------------------------------------------------

%%%%%%%%%%%%%%%%%%%%%%%%%%%%%%%%%%%%%%%%%%%%%%%%%%%%%%%%%
\subsection{Extraction of Moir\'e Parameters from TFM}\label{sup_sec:TFM}

\begin{figure*}[h]
\centering
\includegraphics{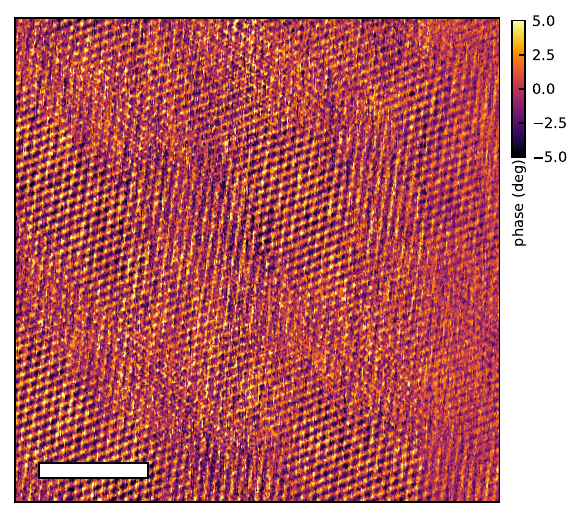}
\caption{\textbf{TFM of HTG.} Full scan range of Fig.~\ref{fig:schematic}D. Scale bar is $\SI{100}{nm}$.
}
\label{sup_fig:full_tfm}
\end{figure*}

The supermoir\'e pattern would be easily discernible if both moir\'e patterns could be imaged simultaneously in HTG. 
Despite the ability to measure subsurface moir\'es in other van der Waals heterostructures using TFM~\cite{pendharkar_torsional_2024}, our TFM images of HTG do not obviously exhibit multiple interfering moir\'e patterns (Fig.~\ref{sup_fig:full_tfm}). 
However, we do observe a slow variation in the contrast of the moir\'e pattern that we believe can be attributed to the supermoir\'e (Fig.~\ref{fig:schematic}D). 

\begin{figure*}[h]
\centering
\includegraphics{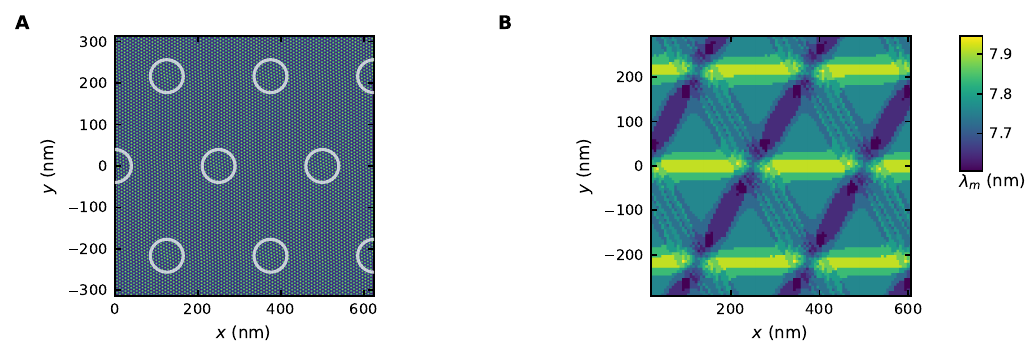}
\caption{\textbf{Visualizing Domain Walls of HTG.} (\textbf{A}) 
Visualizing the stacking region of relaxed HTG.  The bright spots correspond to AA sites in the moir\'e pattern formed by the top two layers of graphene. White circles highlight regions where the top moir\'e pattern is aligned with the bottom moir\'e pattern. (\textbf{B}) Local moir\'e period extracted from $\SI{75}{nm}$ by $\SI{75}{nm}$ windowed FFTs as a function of position in (\textbf{A}). 
}
\label{sup_fig:tfm}
\end{figure*}

In the absence of lattice reconstruction or variations in contrast due to the second moir\'e, an individual moir\'e pattern should not yield any information about the supermoir\'e. 
However, lattice reconstruction favors the formation of moir\'e periodic domains, resulting in a subtle change in the period and rotation of either moir\'e at domain walls relative to the domains~\cite{devakul_magic-angle_2023}. 
In Fig.~\ref{sup_fig:tfm}A, we show the stacking configuration of the top two layers, with the bright regions denoting AA stacking regions.

While this effect is difficult to resolve directly, it can be extracted numerically. To do so, we perform a windowed FFT and raster the window across the image. 
The window size is chosen so that, after applying a Hamming windowing to the truncated data set, roughly five or more periods of the moir\'e remain visible. 
Typically, this corresponds to a window of roughly $\SI{75}{nm}$ by $\SI{75}{nm}$.
This local FFT allows for an extraction of the local moir\'e period, $\lambda_m$, and orientation.

Although the distortion at the domain wall is $\sim1-3$ moir\'e periods, it is clearly resolved as a change in the local moir\'e period as extracted via the windowed FFT (Fig.~\ref{sup_fig:tfm}B). The change in a specific moir\'e wavevector depends on the domain wall's relative orientation. In Fig.~\ref{sup_fig:tfm}, for each window, we extract the wavevector closest to an initial guess of $(2\pi/\sqrt{3}\lambda_m,-2\pi/\lambda_m)$ for $\lambda_m = \SI{7.8}{nm}$. The domain wall that is along this wavevector is less clearly resolved than other directions.

When imaging a real sample using TFM (Fig.~\ref{fig:schematic}D-E), an additional layer of complexity arises because TFM measures the local dynamical friction, not a direct measure of the local stacking. Given that the stripes of reduced moir\'e contrast shown in Fig.~\ref{fig:schematic}E are not observed in samples consisting of a single moir\'e pattern and are highly repeatable, we believe they are likely due to the domain walls given their pitch and spatial extent. As discussed in Sec.~\ref{sup_sec:twist}, the supermoir\'e structure is extremely sensitive to strain. Experimentally realistic amounts of strain can yield a high aspect ratio supermoir\'e similar to what we observe in Fig.~\ref{fig:schematic}D-E.

% ---------------------------------------------------------------------------
% ---------------------------------------------------------------------------

% ---------------------------------------------------------------------------
% ---------------------------------------------------------------------------

%%%%%%%%%%%%%%%%%%%%%%%%%%%%%%%%%%%%%%%%%%%%%%%%%%%%%%%%%
\subsection{Possible Chern Numbers}\label{sup_sec:chern}

In this section, we discuss the possible Chern numbers that can be consistent with a global gap in HTG (and hence quantized transport). We restrict to insulating states that preserve spin-valley $U(1)$ symmetries (i.e.~have definite occupations for the $K\uparrow,K\downarrow,K'\uparrow,K'\downarrow$ flavor sectors), and have an integer partial filling factor within each flavor sector. We also consider the limit of zero/weak magnetic fields such that the notion of a central valence and conduction band in each spin/valley flavor still holds. 

Recall that opening a global gap requires that the spin $s$ and valley $\tau$ resolved Chern numbers of all the occupied bands (including remote valence states that are far from the Fermi level) match across the h-HTG and $\bar{\text{h}}$-HTG domains. In other words, we require $C^\text{h}_{\tau,s}=C^{\bar{\text{h}}}_{\tau,s}$, where $C^d_{\tau,s}$ denotes the total Chern number in domain $d$, valley $\tau$, and spin $s$. Otherwise, there will be topologically enforced gapless edge modes at the domain walls for the flavor sectors where $C^\text{h}_{\tau,s}\neq C^{\bar{\text{h}}}_{\tau,s}$.

If the central bands are empty or full, we have the following Chern numbers of the corresponding gaps~\cite{kwan2024strong,nakatsuji2023multiscale} (we do not consider Landau level gaps on top of these that fundamentally require a magnetic field)
\begin{gather}
    C^{\text{h,empty}}_{K,s}=+1/2,\quad C^{\text{h,full}}_{K,s}=-1/2\\
    C^{\text{h,empty}}_{K',s}=-1/2,\quad C^{\text{h,full}}_{K',s}=+1/2\\
    C^{\bar{\text{h}},\text{empty}}_{K,s}=-1/2,\quad C^{\bar{\text{h}},\text{full}}_{K,s}=+1/2\\
    C^{\bar{\text{h}},\text{empty}}_{K',s}=+1/2,\quad C^{\bar{\text{h}},\text{full}}_{K',s}=-1/2.
\end{gather}
`Empty' leads to a partial filling $\nu^d_{\tau,s}=-1$, while `full' corresponds to $\nu^d_{\tau,s}=+1$.

If a flavor has a single central band filled, then we use $C^{d,\text{v}}_{\tau,s}$ to denote the Chern number of that filled valence central band (this is not the same as the Chern number of the gap above this band). In this case, the total Chern number in that flavor would be $C^{d}_{\tau,s}=C^{\text{d,empty}}_{\tau,s}+C^{d,\text{v}}_{\tau,s}$. This would lead to $\nu^d_{\tau,s}=0$. The total Chern number within a domain is $C^d=\sum_{\tau,s}C^{d}_{\tau,s}$. The total filling within a domain is $\nu^d=\sum_{\tau,s}\nu^d_{\tau,s}$, and must be equal for the two domains for a valid configuration.
$C_{2z}$ symmetry maps the Chern number and filling in $(d,\tau,s)$ to $(\bar{d},\bar{\tau},s)$. We note that cases where the $\nu^d_{\tau,s}$ are not all equal to each other would require the presence of strong interactions or strong external fields. 

While $C^{d,\text{v}}_{\tau,s}$ could in principle take any integer value, we mostly consider the following `strong-coupling' possibilities
\begin{gather}\label{smeq:strongcoupling}
    C^{\text{h,v}}_{K,s}=+1,-2,\quad C^{\text{h,v}}_{K',s}=-1,+2,\quad C^{\bar{\text{h}},\text{v}}_{K,s}=-1,+2\quad C^{\bar{\text{h}},\text{v}}_{K',s}=+1,-2.
\end{gather}
These values are the Chern numbers of the sublattice-Chern basis, which is obtained by diagonalizing the microscopic sublattice operator within the central bands~\cite{devakul_magic-angle_2023}.
This basis enters the strong-coupling theory of HTG, which has been used to predict strongly interacting topological states at integer fillings~\cite{kwan2024strong}. These strong-coupling Chern numbers also coincide with the Chern numbers of the non-interacting orbital Zeeman model for low magnetic fields (see Sec.~\ref{sup_sec:gk_calcs}). Hence, we believe that the strong-coupling Chern numbers capture a wide range of reasonable possibilities for the central band Chern numbers, regardless of the strength of interactions.

In the following, we determine via exhaustive enumeration which possibilities satisfy the global gap condition $C^\text{h}_{\tau,s}=C^{\bar{\text{h}}}_{\tau,s}$ under certain constraints.

% ---------------------------------------------------------------------------
\subsubsection{\texorpdfstring{$C_{2z}$}{C2z} symmetry, equal flavor filling, strong-coupling Chern numbers}
Here, we impose $C_{2z}$ symmetry between the two domains, require equal partial fillings $\nu^\text{h}_{\tau,s}=\nu^{\bar{\text{h}}}_{\tau,s}$, and restrict to the strong-coupling Chern numbers in Eq.~\ref{smeq:strongcoupling}. The only configurations consistent with a global gap occur at charge neutrality $\nu=0$. In fact, $\nu^d_{\tau,s}=0$ is required since the empty-filling (and full-filling) Chern numbers are not compatible across the two domains. The possible Chern numbers are $C=0,\pm6$. The $C=-6$ configuration corresponds to $C^{\text{h,v}}_{K,s}=-2$ and $C^{\text{h,v}}_{K',s}=-1$ (the values for $\bar{\text{h}}$-HTG are fixed by $C_{2z}$), consistent with the orbital Zeeman model. An example configuration for $C=0$ is   $C^{\text{h,v}}_{K,\uparrow}=-2,C^{\text{h,v}}_{K,\downarrow}=+1,C^{\text{h,v}}_{K',\uparrow}=-1,C^{\text{h,v}}_{K',\downarrow}=+2$, which has different Chern numbers in the two spin sectors.

% ---------------------------------------------------------------------------
\subsubsection{\texorpdfstring{$C_{2z}$}{C2z} symmetry, strong-coupling Chern numbers}
Here, we impose $C_{2z}$ symmetry between the two domains, and restrict to the strong-coupling Chern numbers. We do not impose equal partial fillings between the two domains. Again, the only configurations consistent with a global gap occur at charge neutrality $\nu=0$. For $|\nu|=3,4$, we provide a more general argument later why there cannot be a globally gapped state at these fillings, even in the absence of any constraints. For $\nu=-2$, to match say $C^\text{h}_{K,\uparrow}=C^{\bar{\text{h}}}_{K,\uparrow}$, we have two possibilities. Either we occupy both central bands in $(\text{h},K,\uparrow)$ and $(\bar{\text{h}},K',\uparrow)$, but this leaves a Chern number mismatch in the $s=\downarrow$ sectors. Even allowing a further band to be occupied at $\nu=-1$ cannot remedy the mismatch. Or, we occupy the $|C^{d,\text{v}}_{\tau,\uparrow}|>0$ bands in the $s=\uparrow$ sectors, but this again leaves a Chern number mismatch in the $s=\downarrow$ sectors, which again cannot be fixed by occupying an additional band at $\nu=-1$.

At $\nu=0$, the possible Chern numbers for a globally gapped state are $C=0,\pm2,\pm4,\pm6$. For $C=+2$, an example configuration in h-HTG corresponds to emptying the central bands in valley $K$ and fully filling the central bands in valley $K'$. For $C=+4$, an example configuration in h-HTG corresponds to emptying the central bands in $K\uparrow$, filling the central bands in $K'\uparrow$, and occupying the bands with $C^{\text{h,v}}_{K,\downarrow}=+1$ and $C^{\text{h,v}}_{K',\downarrow}=+2$ in the $s=\downarrow$ sector. The $C=+6$ state is obtained by occupying the $|C^{d,\text{v}}_{\tau,s}|>0$ band in each flavor sector.

% ---------------------------------------------------------------------------
\subsubsection{\texorpdfstring{$C_{2z}$}{C2z} symmetry, equal flavor filling}

Here, we impose $C_{2z}$ symmetry between the two domains and require equal partial fillings $\nu^\text{h}_{\tau,s}=\nu^{\bar{\text{h}}}_{\tau,s}$. However, we do not restrict to the strong-coupling Chern numbers in Eq.~\ref{smeq:strongcoupling}. The constraint $\nu^\text{h}_{\tau,s}=\nu^{\bar{\text{h}}}_{\tau,s}$ forces $\nu^d_{\tau,s}=0$ (and hence $\nu=0$) for a globally gapped state. In this case, we can show that any even Chern number $C$ is in principle possible. We can freely choose integer values of $C^\text{h,v}_{K\uparrow}$  and $C^\text{h,v}_{K\downarrow}$, which uniquely fixes the other $C^{d,\text{v}}_{\tau,s}$ using the constraints. The resulting total Chern number is $C=2(C^\text{h,v}_{K\uparrow}+C^\text{h,v}_{K\downarrow}+1)$.

% ---------------------------------------------------------------------------
\subsubsection{Equal flavor filling}
Here, we only impose equal partial fillings $\nu^\text{h}_{\tau,s}=\nu^{\bar{\text{h}}}_{\tau,s}$. The constraint $\nu^\text{h}_{\tau,s}=\nu^{\bar{\text{h}}}_{\tau,s}$ forces $\nu^d_{\tau,s}=0$ (and hence $\nu=0$) for a globally gapped state. Any Chern number $C$ is possible. This is because we can freely choose $C^\text{h,v}_{\tau,s}$, which uniquely fixes $C^{\bar{\text{h}},\text{v}}_{\tau,s}$.

% ---------------------------------------------------------------------------
\subsubsection{No constraints}\label{subsec:noconstraints}
Here, we do not impose any constraints (apart from a fixed global filling factor $\nu^\text{h}=\nu^{\bar{\text{h}}}$). For $|\nu|=3,4$, it is not possible to have a globally gapped state. For $|\nu|=3$, this can be seen by first noting that the system at $|\nu|=4$ has a Chern number mismatch in all four flavor sectors. At say $\nu=-3$ where we are allowed to occupy one central band in each domain, we can make only two flavors have the same Chern number across the domains. An analogous argument holds for $\nu=+3$.

For $|\nu|=2$, it is possible to have a globally gapped state with $C=0,\pm2$. For $|\nu|=0,1$, it is possible to construct a globally gapped state with any total Chern number.

%%%%%%%%%%%%%%%%%%%%%%%%%%%%%%%%%%%%%%%%%%%%%%%%%%%%%%%%%
\subsection{Temperature Dependence, Symmetrization, and Lower Temperature Landau Fans} \label{sup_sec:sym}

\begin{figure*}[h]
\centering
\includegraphics{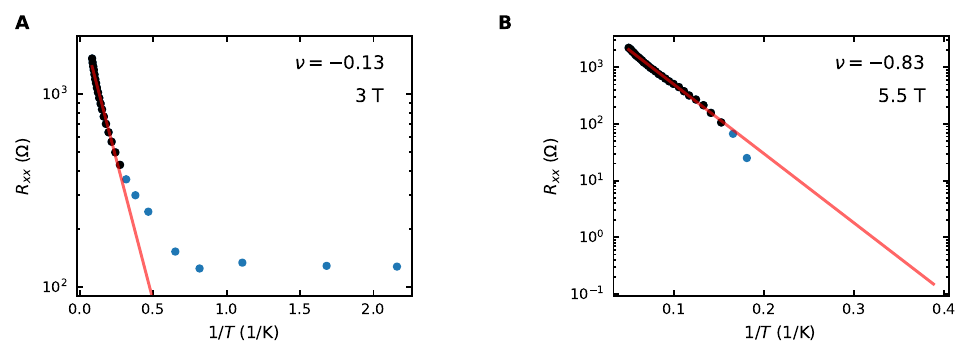}
\caption{\textbf{Arrhenius Activation.} Fit to thermal activation for the $C=-6$ state at (\textbf{A}) $\SI{3}{T}$ and  (\textbf{B}) $\SI{5.5}{T}$. The densities at which the fits are performed correspond to vertical dashed red lines in Fig.~\ref{fig:high_temp_transport} D and E, respectively.
As the temperature is lowered, the resistance begins to deviate from activated behavior. Therefore, blue points are omitted from the fits. 
}
\label{sup_fig:arrhenius}
\end{figure*}

We fit the temperature dependence of the minimum in the longitudinal resistance at $\SI{3}{T}$ and $\SI{5.5}{T}$ (dashed red lines in Fig.~\ref{fig:high_temp_transport}D-E, respectively) to the Arrhenius equation, $R_{xx} = R_0 \exp\{-\Delta/{2k_\mathrm{B}T}\}$, where $\Delta$ is the thermal activation gap and $k_\mathrm{B}$ is Boltzmann's constant. 
The fit to the high temperature points (denoted in black) yields gaps of $1.17\pm 0.04$\,meV and $4.86\pm 0.05$\,meV, respectively (Fig.~\ref{sup_fig:arrhenius}). 
For temperatures below $\SI{5}{K}$, $R_{xx}<0$ at $\SI{5.5}{T}$, presumably due to mixing.  

Various phenomena can lead to a mixing of voltage signals in typical Hall bar measurements (i.e.~introducing longitudinal character into a Hall measurement or vice versa), including geometric shifts between probes, anisotropic conduction, or inhomogeneity within a device. 
To attempt to remove the effects of mixing, a standard approach is to (anti)symmetrize the longitudinal resistivity (Hall resistance) as a function of magnetic field. 

\begin{figure*}[h]
\centering
\includegraphics{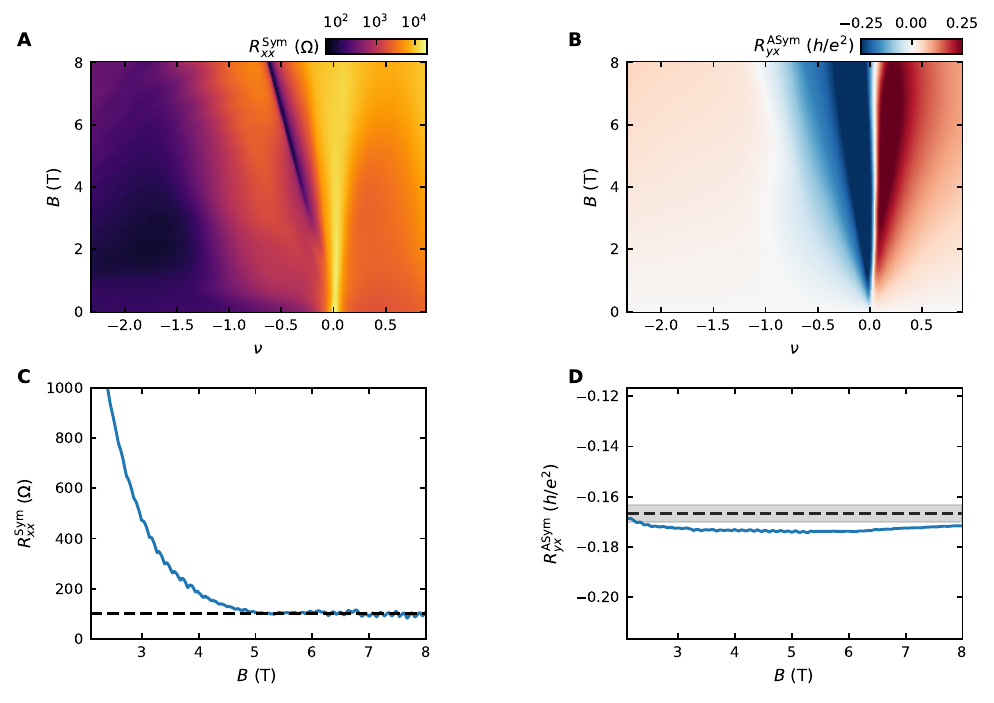}
\caption{\textbf{Symmetrized Magnetotransport.} (\textbf{A}) Symmetrized longitudinal resistance and (\textbf{B}) antisymmetrized Hall resistance as a function of carriers per moir\'e unit cell $\nu$ and magnetic field $B$. 
(\textbf{C}) Symmetrized longitudinal resistance and (\textbf{D}) antisymmetrized Hall resistance along the $C=-6$ gap. In (\textbf{C}), the dashed horizontal line corresponds to $\SI{100}{\Omega}$. In (\textbf{D}), the dashed horizontal line corresponds to $-h/6e^2$ with a $\pm2$\% window shaded in gray. All data are taken at $\SI{4.6}{K}$.
}
\label{sup_fig:sym}
\end{figure*}

As shown in Fig.~\ref{sup_fig:sym}, the longitudinal resistance is minimally changed by symmetrization: comparing to the data shown in Fig.~\ref{fig:high_temp_transport}, the raw and symmetrized data differ by $<\SI{20}{\Omega}$ for fields above $\SI{3}{T}$. 
The antisymmetrized Hall resistance is $60-\SI{125}{\Omega}$ larger than the raw data for fields above $\SI{3}{T}$, which corresponds to 1.4\% to 2.9\% deviations from quantization. 
This amount of deviation from quantization is within the error expected from drift in the gain in our preamplifier and demodulation~\cite{fischer_influence_2005}. 
The observable sawtooth-like behavior in the presented line cuts comes from the fact that we are interpolating across a finite resolution two-dimensional data set; quantization could be slightly better if density were more optimally tuned for each field.

\begin{figure*}[h]
\centering
\includegraphics{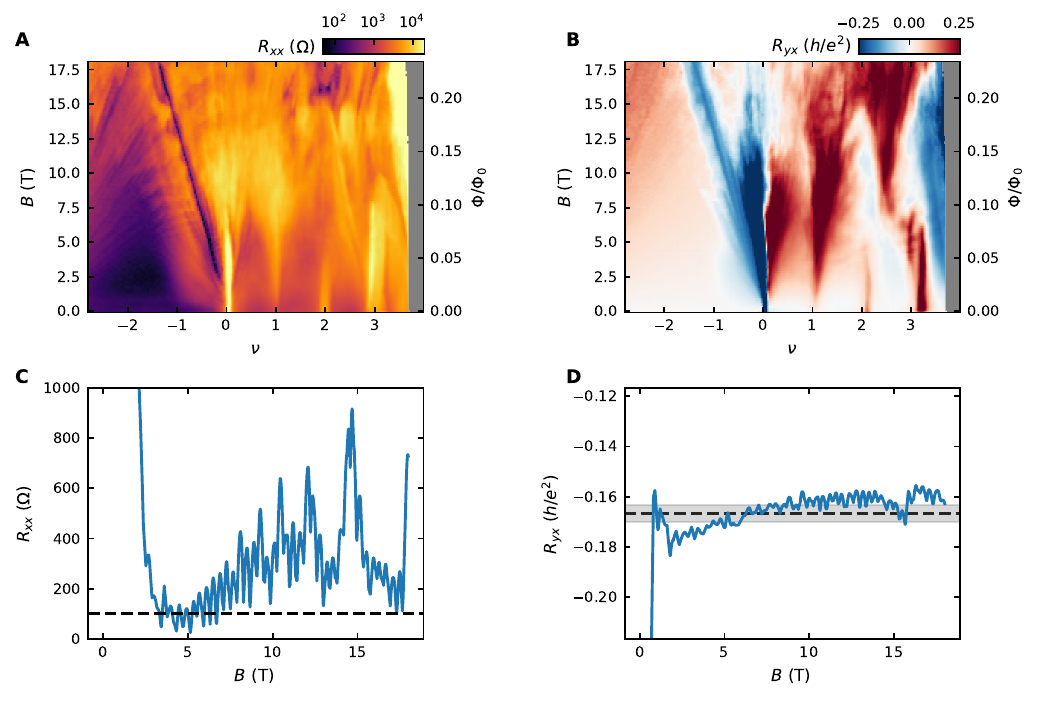}
\caption{\textbf{High Field Landau Fan.} 
(\textbf{A}) Longitudinal resistance and
(\textbf{B}) Hall resistance as a function of field  up to $\SI{18}{T}$  and carriers per moir\'e unit cell at zero displacement field and $\SI{300}{mK}$.
(\textbf{C}) Line cut of longitudinal resistance and (\textbf{D}) Hall resistance along the $C=-6$ gap. 
In (\textbf{C}), the dashed horizontal line corresponds to $\SI{100}{\Omega}$. In (\textbf{D}), the dashed horizontal line corresponds to $-h/6e^2$ with a $\pm2$\% window shaded in gray. 
Spikes of increased longitudinal resistance are likely artifacts of insufficient resolution to accurately track the minimum in resistance along this gap as a function of field.
}
\label{sup_fig:maglabfan}
\end{figure*}

\begin{figure*}[h]
\centering
\includegraphics{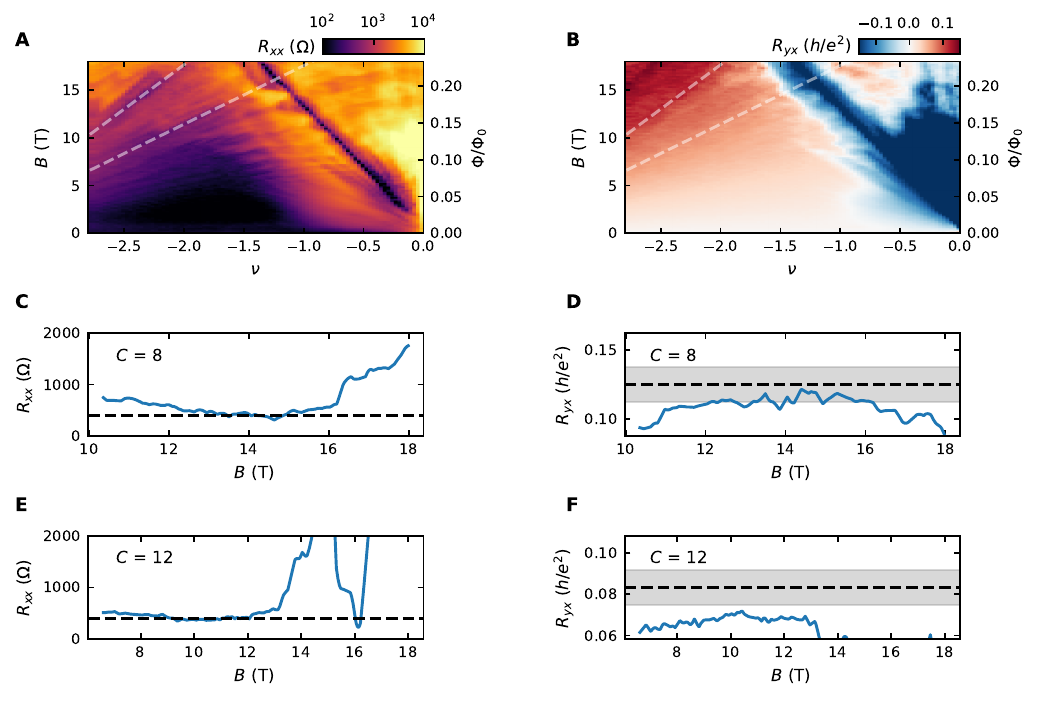}
\caption{\textbf{Landau Levels from Full Filling.} (\textbf{A}) Longitudinal resistance and (\textbf{B}) Hall resistance as a function of carriers per moir\'e unit cell $\nu$ and magnetic field $B$ reproduced from Fig.~\ref{sup_fig:maglabfan} and zoomed near $\nu=-4$. 
Faint dashed white lines correspond to $C=8$ and $12$ St\v{r}eda lines emanating from $\nu=-4$. 
(\textbf{C}) Longitudinal resistance and (\textbf{D}) Hall resistance along the drawn $C=8$ line. 
(\textbf{E}) Longitudinal resistance and (\textbf{F}) Hall resistance along the drawn $C=12$ line. 
In (\textbf{C}) and (\textbf{E}), the minimum longitudinal resistance reached along either St\v{r}eda line is approximately $\SI{400}{\Omega}$ (excluding points after the longitudinal resistance begins to rise).
The black dashed lines in (\textbf{D}) and (\textbf{F}) correspond to $h/Ce^2$, with the gray shaded region indicating a window of $\pm10\%$ about perfect quantization.
}
\label{sup_fig:minus4_fan}
\end{figure*}

Upon cooling to $\SI{300}{mK}$, we begin to see the appearance of modulations of the resistance adjacent to the $C=-6$ gap at higher magnetic fields (Fig.~\ref{sup_fig:maglabfan}A). 
These modulations appear near simple fractions of magnetic flux per moir\'e unit cell and are likely due to the non-monotonicity of the gap with field when in the Hofstadter regime. 
This effect is potentially exacerbated by disorder in the sample. 
Given the resolution of this data set, oscillations in the line cut of $R_{xx}$ become much more pronounced (Figs.~\ref{sup_fig:maglabfan}C and D). 
We see that the longitudinal resistance drops below the approximate minimum value of $\SI{100}{\Omega}$ seen at $\SI{4.6}{K}$. 
In the Hall resistance, we observe a similar level of quantization with deviations from the higher temperature data likely dominated by mixing.

\begin{figure*}[h]
\centering
\includegraphics{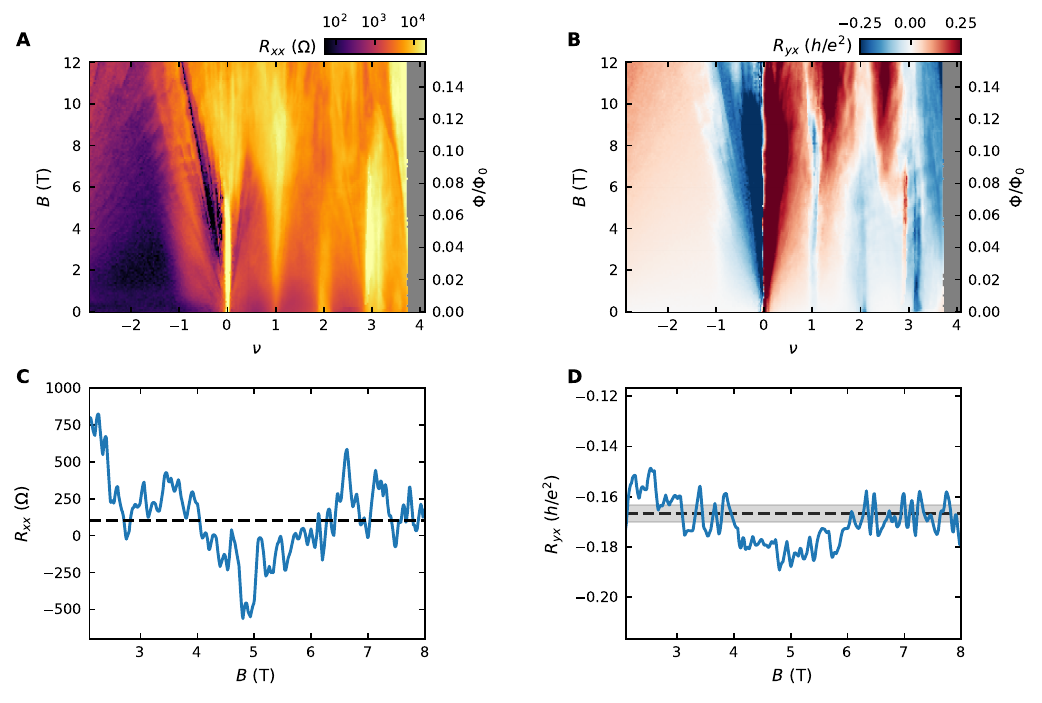}
\caption{\textbf{Landau Fan at Dilution Refrigerator Temperatures.} 
(\textbf{A}) Longitudinal and 
(\textbf{B}) Hall resistance as a function of magnetic field and carriers per moir\'e unit cell at $\SI{37}{mK}$.
(\textbf{C}) Line cut of longitudinal resistance and (\textbf{D}) Hall resistance along the $C=-6$ gap. In (\textbf{C}), the dashed horizontal line corresponds to $\SI{100}{\Omega}$. In (\textbf{D}), the dashed horizontal line corresponds to $-h/6e^2$ with a $\pm2$\% window shaded in gray.
The longitudinal resistance dropping substantially below zero is indicative of substantial mixing. 
Data in (\textbf{C}) and (\textbf{D}) are from a separate higher resolution fan than (\textbf{A}) and (\textbf{B}) that only extends up to 8 T.
}
\label{sup_fig:dilfan}
\end{figure*}

\begin{figure*}[h]
\centering
\includegraphics{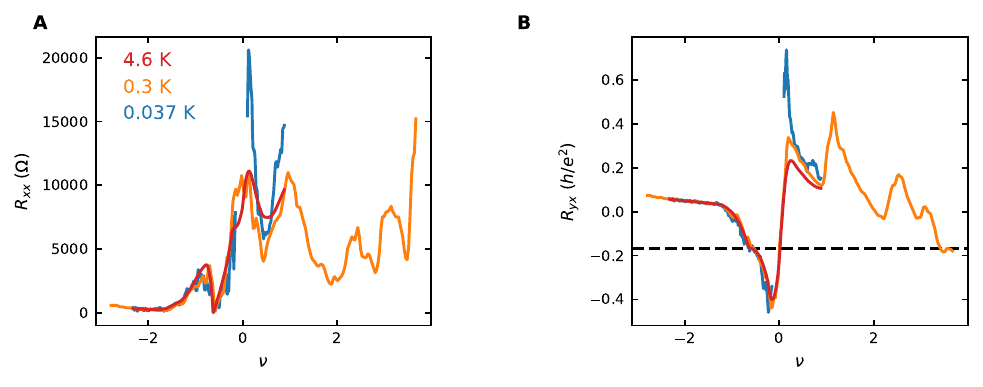}
\caption{\textbf{Linecuts of Landau Fans at 8 T.} 
Linecuts of (\textbf{A}) Longitudinal and 
(\textbf{B}) Hall resistance at $\SI{8}{T}$. 
Red, orange, and blue curves correspond to cuts of the $\SI{4.6}{K}$ (Fig.~\ref{fig:high_temp_transport}), $\SI{300}{mK}$ (Fig.~\ref{sup_fig:maglabfan}), and $\SI{37}{mK}$ (Fig.~\ref{sup_fig:dilfan}) fans, respectively. In (\textbf{B}), the dashed horizontal line corresponds to $-h/6e^2$.
}
\label{sup_fig:alltemps_cut}
\end{figure*}

Cooling further to dilution refrigerator temperatures ($\SI{37}{mK}$), we observe modest qualitative changes in the Landau fan of longitudinal resistance; several minima in longitudinal resistance near the $C=-6$ gap become more pronounced (Fig.~\ref{sup_fig:dilfan}A).
We do, however, observe a fairly dramatic change in the Landau fan of Hall resistance (Fig.~\ref{sup_fig:dilfan}B). We measure a change in the sign of the Hall resistance near $\nu=1$, $2$, and $3$. 
While the most likely explanation is simply the reduced temperature, we cannot explicitly rule out a change in the supermoir\'e structure from thermal cycling~\cite{hoke_imaging_2024}.
Taking a line cut along the $C=-6$ gap, we see that the longitudinal resistance goes negative, likely through some combination of mixing and disorder in the device (Fig.~\ref{sup_fig:dilfan}C). 
A line cut of the Hall resistance exhibits quantization comparable to the $\SI{300}{mK}$ data over the appropriate field range, corroborating the conclusion that quantization is not limited by thermal broadening (Fig.~\ref{sup_fig:dilfan}D).

% ---------------------------------------------------------------------------
% ---------------------------------------------------------------------------

% ---------------------------------------------------------------------------
% ---------------------------------------------------------------------------

% ---------------------------------------------------------------------------
% ---------------------------------------------------------------------------

% ---------------------------------------------------------------------------
% ---------------------------------------------------------------------------

% ---------------------------------------------------------------------------
% ---------------------------------------------------------------------------

% ---------------------------------------------------------------------------
% ---------------------------------------------------------------------------

%%%%%%%%%%%%%%%%%%%%%%%%%%%%%%%%%%%%%%%%%%%%%%%%%%%%%%%%%
\subsection{Displacement Field Dependence} \label{sup_sec:dfield}

\begin{figure*}[h]
\centering
\includegraphics{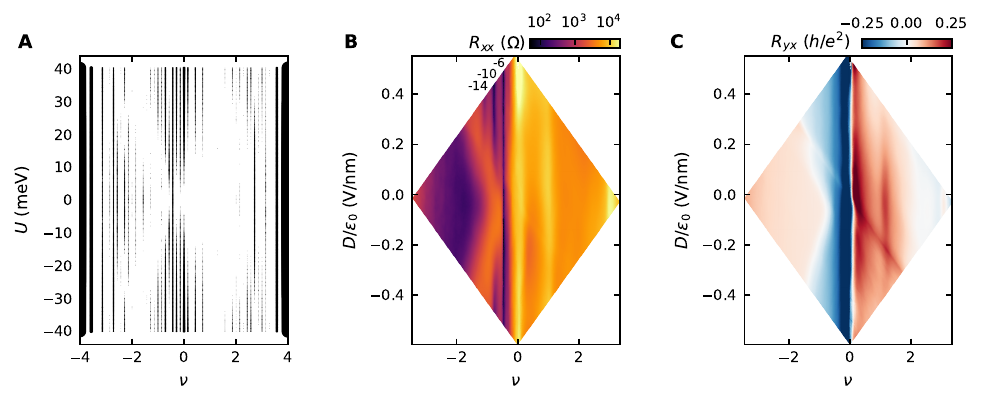}
\caption{\textbf{Effect of Displacement Field.} 
(\textbf{A}) Calculated gap sizes in the Hofstadter spectrum as a function of interlayer potential energy $U$ and carriers per moir\'e unit cell at a fixed field of $\SI{5.85}{T}$. The size of individual dots corresponds to the size of the gap in the spectrum.
(\textbf{B}) Gate map of longitudinal resistance $R_{xx}$ and (\textbf{C}) Hall resistance $R_{yx}$ at \SI{-6}{T} and \SI{4.6}{K}.
}
\label{sup_fig:nD}
\end{figure*}

The gate map in finite magnetic field demonstrates that the $C=-6$ gap is robust over our full accessible range of displacement field (Fig.~\ref{sup_fig:nD}B and C).
For further hole doping, we see a `$>$'-shaped feature of increased resistance that was previously identified as a Van Hove singularity~\cite{xia_topological_2025}. 
For moderate displacement fields, we also measure strong dips in resistance at fillings that would correspond to $C = -10$ and $-14$. 
In Fig.~\ref{sup_fig:nD}A, we show the gap sizes expected from our single-particle Hofstadter calculations within an h-HTG domain as a function of $\nu$ and interlayer potential $U$. 
As the conduction band exhibits strong correlations in transport measurements, we do not expect our single-particle calculation to capture the experimental phenomenology for electron doping. 
For hole doping, we find a similar development of a sequence of gaps at moderate $U$.

\begin{figure*}[h]
\centering
\includegraphics{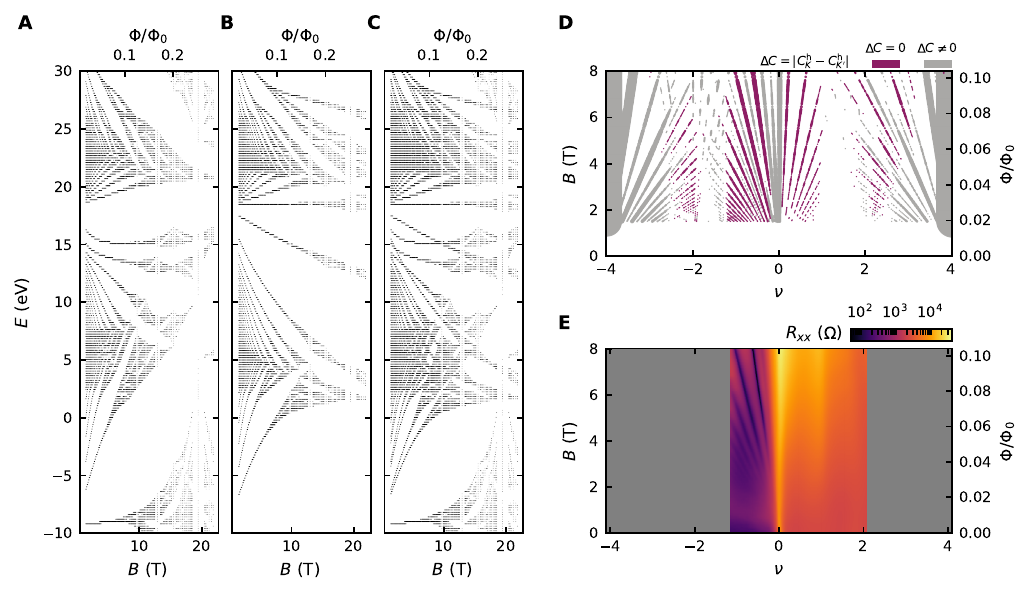}
\caption{\textbf{Hofstadter Spectra and Wannier Diagram at Moderate Displacement Field.} 
Calculated Hofstadter spectrum for (\textbf{A}) $K$ in h-HTG (equivalently $K'$ in $\bar{\text{h}}$-HTG), (\textbf{B}) $K'$ in h-HTG (equivalently $K$ in $\bar{\text{h}}$-HTG), and (\textbf{C}) both $K$ and $K'$ valleys combined for an applied interlayer potential of $U=\SI{25}{meV}$. 
(\textbf{D}) Associated Wannier diagram for (\textbf{C}). The size of individual dots corresponds to the size of the gap in the spectrum. 
(\textbf{E}) Longitudinal resistance $R_{xx}$ as a function of carriers per moir\'e unit cell and magnetic field for $D/\epsilon_0=\SI{0.35}{V/nm}$ and $T=\SI{4.6}{K}$.}
\label{sup_fig:highD_fan}
\end{figure*}

A Landau fan plot confirms that these states emanate from the CNP (Fig.~\ref{sup_fig:highD_fan}).
The Wannier diagram of the Hofstadter calculation shows a full sequence of even Chern number gaps emanating from the CNP.
However, with the exception of a poorly quantized $C=-4$ state, we experimentally measure a series of $C=-6, -10,  -14, \dots$ gaps (Fig.~\ref{sup_fig:highD_fan_cuts}).
In this scenario, the Chern number associated with the gap in each spin-valley flavor decreases by 1, which preserves equal Chern numbers across h-HTG and $\bar{\text{h}}$-HTG for each flavor sector.
A two-fold sequence is possible in the presence of significant spin-splitting.
We note that $C=-4$ could also arise from a $C_{2z}$-symmetric strong coupling state with substantial flavor polarization (see Sec.~\ref{sup_sec:chern}).

% ---------------------------------------------------------------------------
% ---------------------------------------------------------------------------

% ---------------------------------------------------------------------------
% ---------------------------------------------------------------------------

% ---------------------------------------------------------------------------
\begin{figure*}[h]
\centering
\includegraphics{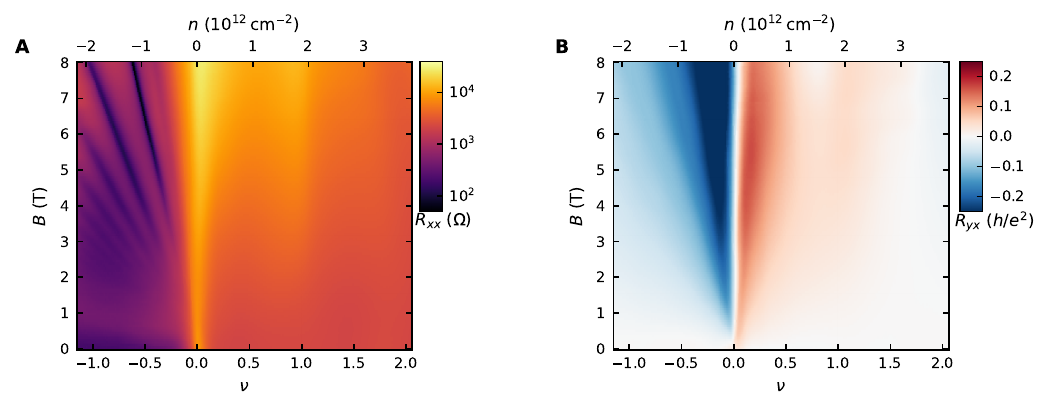}
\caption{\textbf{Landau Fans at Moderate Displacement Field.} 
(\textbf{A}) Longitudinal resistance and 
(\textbf{B}) Hall resistance as a function of carriers per moir\'e unit cell and magnetic field at a displacement field of $D/\epsilon_0=\SI{0.35}{V/nm}$ and $\SI{4.6}{K}$.}
\label{sup_fig:highD_fan_xxxy}
\end{figure*}
% ---------------------------------------------------------------------------

% ---------------------------------------------------------------------------
\begin{figure*}[h]
\centering
\includegraphics{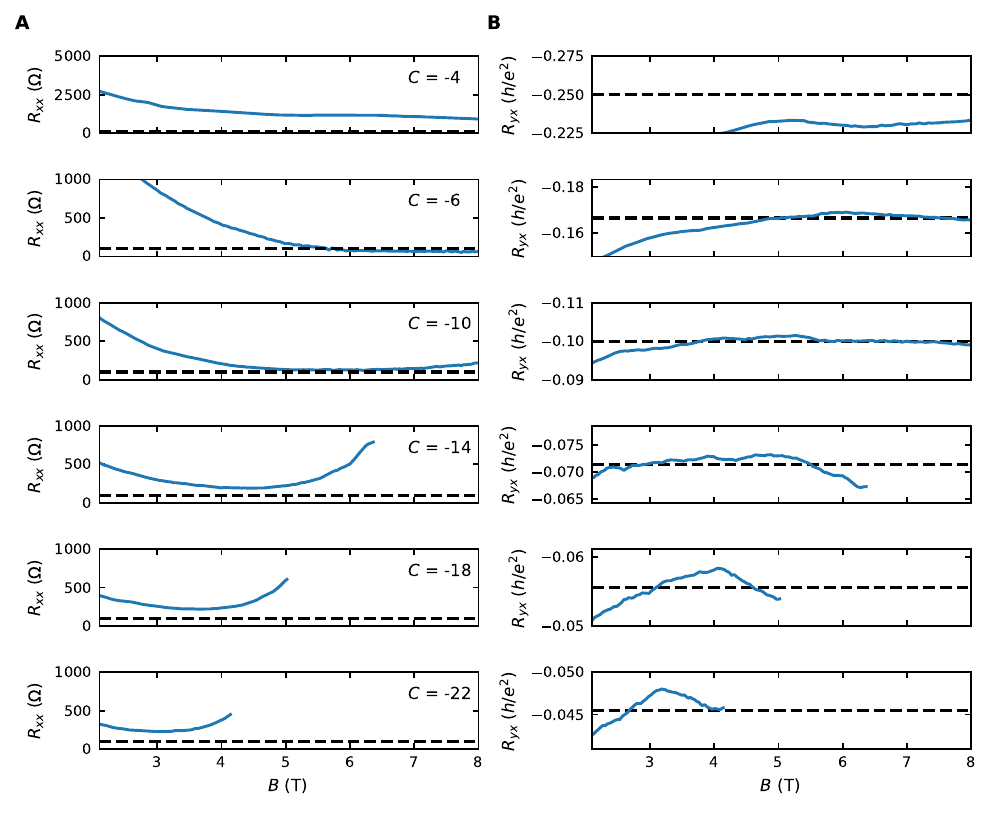}
\caption{\textbf{Cuts of Landau Fan at Moderate Displacement Field.} 
(\textbf{A}) Longitudinal resistance and the concurrent
(\textbf{B}) Hall resistance along various St\v{r}eda lines of Fig.~\ref{sup_fig:highD_fan_xxxy} as indicated for the Chern number shown in the subpanel of (\textbf{A}). 
In (\textbf{A}), the horizontal dashed lines indicate $\SI{100}{\Omega}$. 
In (\textbf{B}), the horizontal dashed lines indicate the expected quantized value $h/Ce^2$ with a $\pm2\%$ window around this value shaded in gray.
The accessible extent for higher-$C$ line cuts is limited by accessible gate range at this displacement field of $D/\epsilon_0=\SI{0.35}{V/nm}$. Data are taken at $\SI{4.6}{K}$.}
\label{sup_fig:highD_fan_cuts}
\end{figure*}
% ---------------------------------------------------------------------------

%%%%%%%%%%%%%%%%%%%%%%%%%%%%%%%%%%%%%%%%%%%%%%%%%%%%%%%%%
\subsection{Angle Dependence of Hofstadter}\label{sup_sec:hof}

\begin{figure*}[h]
\centering
\includegraphics{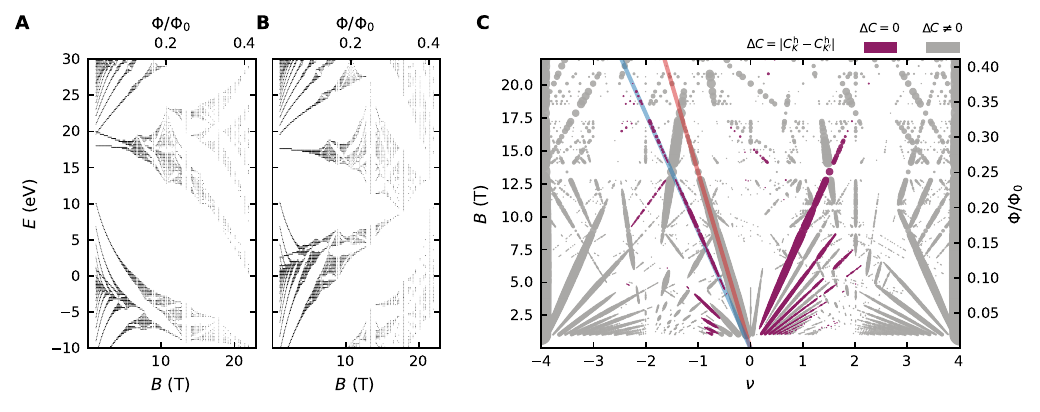}
\caption{\textbf{Hofstadter Spectrum and Wannier diagram for 1.49\textdegree\,HTG.} Calculated Hofstadter spectrum for (\textbf{A}) $K$ in h-HTG (equivalently $K'$ in $\bar{\text{h}}$-HTG), (\textbf{B}) $K'$ in h-HTG (equivalently $K$ in $\bar{\text{h}}$-HTG).
(\textbf{C}) Associated Wannier diagram for (\textbf{A}) and (\textbf{B}) combined.  
Fillings $\nu$ and assigned Chern numbers include a factor of two for spin degeneracy.
The size of each dot corresponds to the size of the gap in the spectrum.
Each gap is color-coded according to Eq.~\ref{eq:global_gap_condition}, evaluated within h-HTG as the difference between its two valleys, $\Delta C = \left|C^{\text{h}}_{K} - C^{\text{h}}_{K'}\right|$.
Magenta dots indicate gaps where h-HTG and $\bar{\text{h}}$-HTG have the same flavor-resolved Chern numbers, and are therefore candidates for global gaps. 
Gray dots correspond to cases where h-HTG and $\bar{\text{h}}$-HTG have differing flavor-resolved Chern numbers and therefore a network of edge modes is expected.
$C=-4$ and $-6$ gaps from the CNP are indicated in red and blue, respectively.
}
\label{sup_fig:1p49_hoff}
\end{figure*}

\begin{figure*}[h]
\centering
\includegraphics{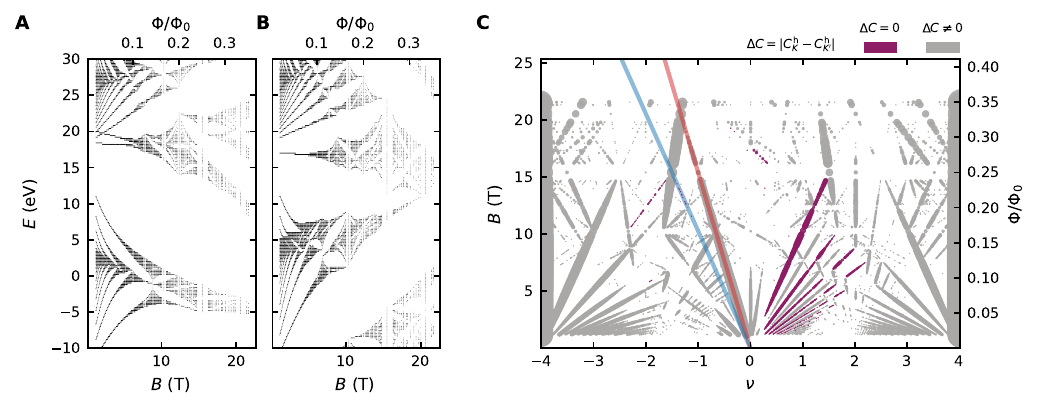}
\caption{\textbf{Hofstadter Spectrum and Wannier diagram for 1.60\textdegree\,HTG.} Calculated Hofstadter spectrum for (\textbf{A}) $K$ in h-HTG (equivalently $K'$ in $\bar{\text{h}}$-HTG), (\textbf{B}) $K'$ in h-HTG (equivalently $K$ in $\bar{\text{h}}$-HTG).
(\textbf{C}) Associated Wannier diagram for (\textbf{A}) and (\textbf{B}) combined. 
Fillings $\nu$ and assigned Chern numbers include a factor of two for spin degeneracy.
The size of each dot corresponds to the size of the gap in the spectrum.
Each gap is color-coded according to Eq.~\ref{eq:global_gap_condition}, evaluated within h-HTG as the difference between its two valleys, $\Delta C = \left|C^{\text{h}}_{K} - C^{\text{h}}_{K'}\right|$.
Magenta dots indicate gaps where h-HTG and $\bar{\text{h}}$-HTG have the same flavor-resolved Chern numbers, and are therefore candidates for global gaps. 
Gray dots correspond to cases where h-HTG and $\bar{\text{h}}$-HTG have differing flavor-resolved Chern numbers and therefore a network of edge modes is expected.
$C=-4$ and $-6$ gaps from the CNP are indicated in red and blue, respectively.
}
\label{sup_fig:1p6_hoff}
\end{figure*}

\begin{figure*}[h]
\centering
\includegraphics{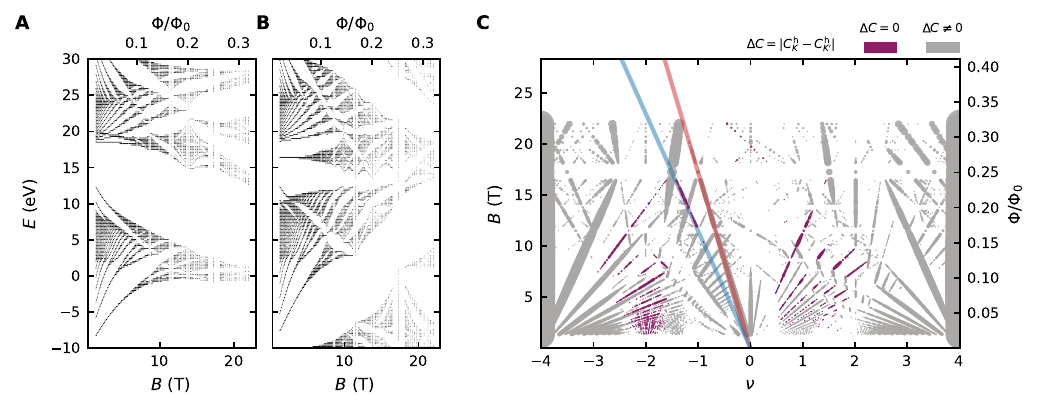}
\caption{\textbf{Hofstadter Spectrum and Wannier diagram for 1.69\textdegree\,HTG.} Calculated Hofstadter spectrum for (\textbf{A}) $K$ in h-HTG (equivalently $K'$ in $\bar{\text{h}}$-HTG), (\textbf{B}) $K'$ in h-HTG (equivalently $K$ in $\bar{\text{h}}$-HTG).
(\textbf{C}) Associated Wannier diagram for (\textbf{A}) and (\textbf{B}) combined. 
Fillings $\nu$ and assigned Chern numbers include a factor of two for spin degeneracy.
The size of each dot corresponds to the size of the gap in the spectrum.
Each gap is color-coded according to Eq.~\ref{eq:global_gap_condition}, evaluated within h-HTG as the difference between its two valleys, $\Delta C = \left|C^{\text{h}}_{K} - C^{\text{h}}_{K'}\right|$.
Magenta dots indicate gaps where h-HTG and $\bar{\text{h}}$-HTG have the same flavor-resolved Chern numbers, and are therefore candidates for global gaps. 
Gray dots correspond to cases where h-HTG and $\bar{\text{h}}$-HTG have differing flavor-resolved Chern numbers and therefore a network of edge modes is expected.
$C=-4$ and $-6$ gaps from the CNP are indicated in red and blue, respectively.
}
\label{sup_fig:1p69_hoff}
\end{figure*}

\begin{figure*}[h]
\centering
\includegraphics{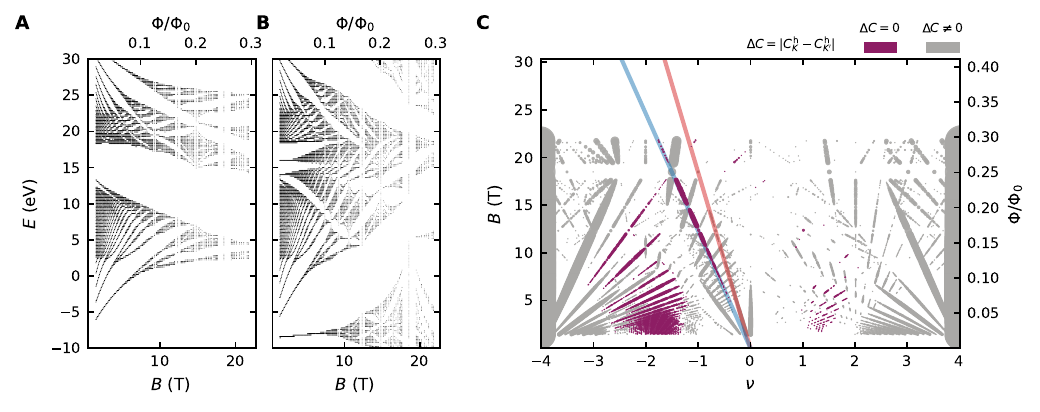}
\caption{\textbf{Hofstadter Spectrum and Wannier diagram for 1.75\textdegree\,HTG.} Calculated Hofstadter spectrum for (\textbf{A}) $K$ in h-HTG (equivalently $K'$ in $\bar{\text{h}}$-HTG), (\textbf{B}) $K'$ in h-HTG (equivalently $K$ in $\bar{\text{h}}$-HTG).
(\textbf{C}) Associated Wannier diagram for (\textbf{A}) and (\textbf{B}) combined. 
Fillings $\nu$ and assigned Chern numbers include a factor of two for spin degeneracy.
The size of each dot corresponds to the size of the gap in the spectrum.
Each gap is color-coded according to Eq.~\ref{eq:global_gap_condition}, evaluated within h-HTG as the difference between its two valleys, $\Delta C = \left|C^{\text{h}}_{K} - C^{\text{h}}_{K'}\right|$.
Magenta dots indicate gaps where h-HTG and $\bar{\text{h}}$-HTG have the same flavor-resolved Chern numbers, and are therefore candidates for global gaps. 
Gray dots correspond to cases where h-HTG and $\bar{\text{h}}$-HTG have differing flavor-resolved Chern numbers and therefore a network of edge modes is expected.
$C=-4$ and $-6$ gaps from the CNP are indicated in red and blue, respectively.
}
\label{sup_fig:1p75_hoff}
\end{figure*}

\begin{figure*}[h]
\centering
\includegraphics{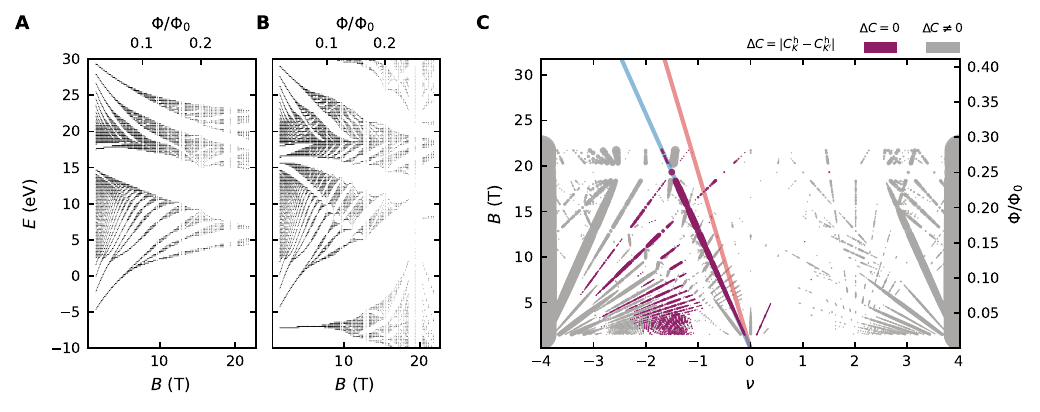}
\caption{\textbf{Hofstadter Spectrum and Wannier diagram for 1.79\textdegree\,HTG.} Calculated Hofstadter spectrum for (\textbf{A}) $K$ in h-HTG (equivalently $K'$ in $\bar{\text{h}}$-HTG), (\textbf{B}) $K'$ in h-HTG (equivalently $K$ in $\bar{\text{h}}$-HTG).
(\textbf{C}) Associated Wannier diagram for (\textbf{A}) and (\textbf{B}) combined. 
Fillings $\nu$ and assigned Chern numbers include a factor of two for spin degeneracy.
The size of each dot corresponds to the size of the gap in the spectrum.
Each gap is color-coded according to Eq.~\ref{eq:global_gap_condition}, evaluated within h-HTG as the difference between its two valleys, $\Delta C = \left|C^{\text{h}}_{K} - C^{\text{h}}_{K'}\right|$.
Magenta dots indicate gaps where h-HTG and $\bar{\text{h}}$-HTG have the same flavor-resolved Chern numbers, and are therefore candidates for global gaps. 
Gray dots correspond to cases where h-HTG and $\bar{\text{h}}$-HTG have differing flavor-resolved Chern numbers and therefore a network of edge modes is expected.
$C=-4$ and $-6$ gaps from the CNP are indicated in red and blue, respectively.
}
\label{sup_fig:1p79_hoff}
\end{figure*}

\begin{figure*}[h]
\centering
\includegraphics{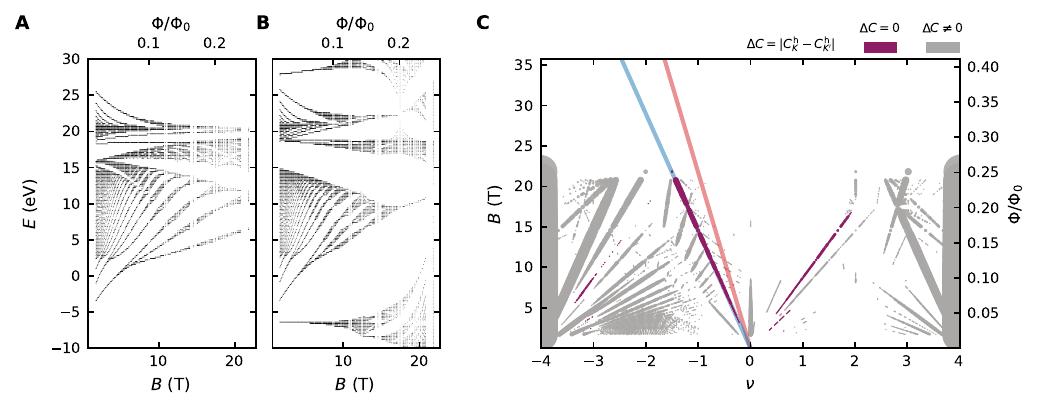}
\caption{\textbf{Hofstadter Spectrum and Wannier diagram for 1.90\textdegree\,HTG.} Calculated Hofstadter spectrum for (\textbf{A}) $K$ in h-HTG (equivalently $K'$ in $\bar{\text{h}}$-HTG), (\textbf{B}) $K'$ in h-HTG (equivalently $K$ in $\bar{\text{h}}$-HTG).
(\textbf{C}) Associated Wannier diagram for (\textbf{A}) and (\textbf{B}) combined. 
Fillings $\nu$ and assigned Chern numbers include a factor of two for spin degeneracy.
The size of each dot corresponds to the size of the gap in the spectrum.
Each gap is color-coded according to Eq.~\ref{eq:global_gap_condition}, evaluated within h-HTG as the difference between its two valleys, $\Delta C = \left|C^{\text{h}}_{K} - C^{\text{h}}_{K'}\right|$.
Magenta dots indicate gaps where h-HTG and $\bar{\text{h}}$-HTG have the same flavor-resolved Chern numbers, and are therefore candidates for global gaps. 
Gray dots correspond to cases where h-HTG and $\bar{\text{h}}$-HTG have differing flavor-resolved Chern numbers and therefore a network of edge modes is expected.
$C=-4$ and $-6$ gaps from the CNP are indicated in red and blue, respectively.
}
\label{sup_fig:1p9_hoff}
\end{figure*}

Here we present a sequence of Hofstadter spectra and their associated Wannier diagrams from $\theta=\SI{1.49}{\degree}$ to $\SI{1.90}{\degree}$ (Figs.~\ref{sup_fig:1p49_hoff}--\ref{sup_fig:1p9_hoff}).
Throughout this section, the Chern number quoted for each valley accounts for twofold spin degeneracy.
To produce the valley-resolved coloring of the combined Wannier diagrams, each gap in the $K$ and $K'$ Hofstadter spectra must be assigned a Chern number $C$. In the Wannier representation, gaps trace lines satisfying the Diophantine equation $\nu = C\phi/\phi_0 + s$, where $s$ is the zero-field filling offset. The TKNN formalism~\cite{TKNN} extracts $C$ from the Berry curvature of the Bloch eigenstates; however, our simulation outputs only the energy spectrum, so a direct topological calculation is not available. Instead, we adopt a gap-tracing approach: candidate St\v{r}eda lines are enumerated for integer $C$ and physically allowed $s$ values. Gaps in the Hofstadter spectrum are then assigned to the best-matching St\v{r}eda line based on the density of nearby gaps along the candidate line that exhibit a gap size of similar magnitude with a preference for principal filling offsets ($s \in 4\mathbb{Z}$). The Chern numbers are assigned independently for the $K$ and $K'$ valley spectra, and the combined Wannier diagram encodes $\Delta C = \left|C_K - C_{K'}\right|$ via the dot color.

We find the Hofstadter gaps for the valence band to generally be larger than those of the conduction band due to the conduction band having a narrower bandwidth at zero magnetic field. 
We note that the most dominant gap at the CNP changes from $C=-4$ at smaller twist angles ($\theta=\SI{1.49}{\degree}$--$\SI{1.69}{\degree}$, Figs.~\ref{sup_fig:1p49_hoff}--\ref{sup_fig:1p69_hoff}) to $C=-6$ at larger twist angles ($\theta=\SI{1.75}{\degree}$--$\SI{1.90}{\degree}$, Figs.~\ref{sup_fig:1p75_hoff}--\ref{sup_fig:1p9_hoff}).
For a $C=-4$ gap, quantization is not expected as such a situation does not have a constant Chern number across domains within each flavor.
The $K'$ valley $-3$ gap remains large throughout this range of twist angles. 
To achieve a total Chern number of $C=-4$, the $K$ valley must be in the $-1$ gap.

% ---------------------------------------------------------------------------

% ---------------------------------------------------------------------------

% ---------------------------------------------------------------------------
% ---------------------------------------------------------------------------

% ---------------------------------------------------------------------------
% ---------------------------------------------------------------------------

% ---------------------------------------------------------------------------
% ---------------------------------------------------------------------------

% ---------------------------------------------------------------------------
% ---------------------------------------------------------------------------

% ---------------------------------------------------------------------------
% ---------------------------------------------------------------------------

%%%%%%%%%%%%%%%%%%%%%%%%%%%%%%%%%%%%%%%%%%%%%%%%%%%%%%%%%
\subsection{Additional Devices} \label{sup_sec:other_devices}

\begin{figure*}[h]
\centering
\includegraphics{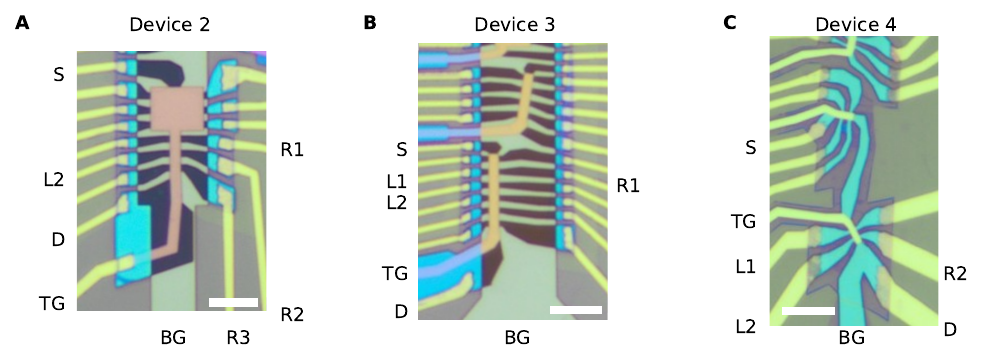}
\caption{\textbf{Additional HTG Devices.}
Optical micrographs of (\textbf{A}) Device 2 and (\textbf{B}) Device 3. The twist angles of these devices, as extracted from densities corresponding to full filling, are $\SI{1.95}{\degree}$ and $\SI{1.85}{\degree}$, respectively. Labels indicate the contact used for current sourcing (S), current draining (\textbf{D}), back gate (BG), top gate (TG), and voltage probes on the left (L) and right (R).
All scale bars are $\SI{5}{\micro\meter}$.
}
\label{sup_fig:device_schematics}
\end{figure*}

Here, we consider transport in two additional devices with twist angles of $\SI{1.95}{\degree}$ and $\SI{1.85}{\degree}$ (Fig.~\ref{sup_fig:device_schematics}).

\begin{figure*}[h]
\centering
\includegraphics{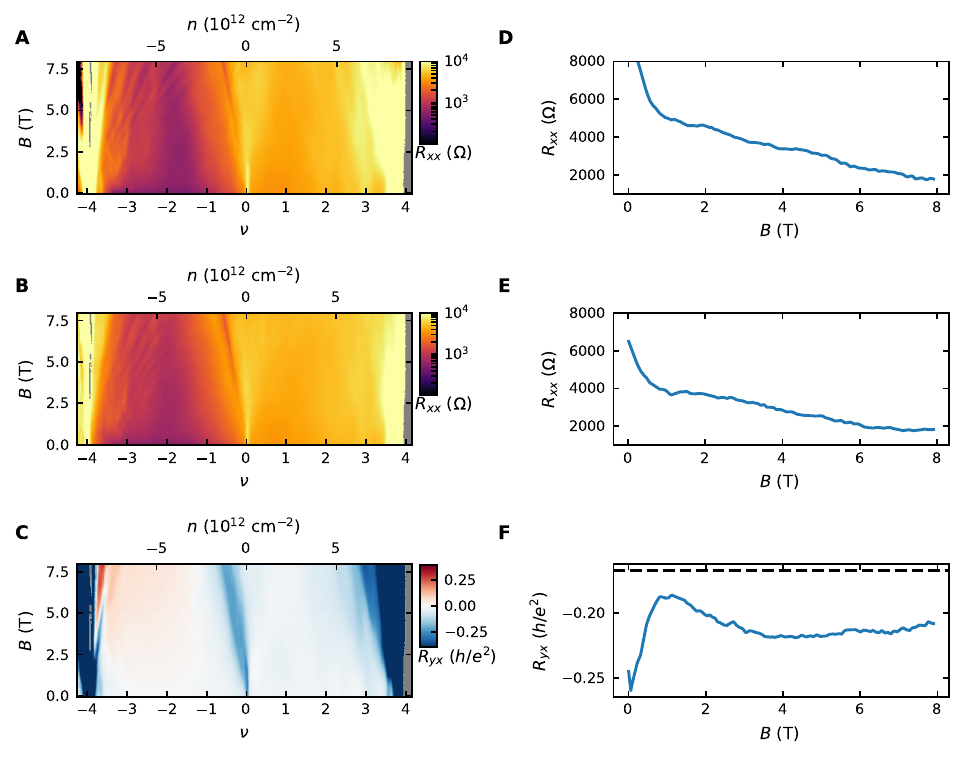}
\caption{\textbf{Landau Fans for Device 2 (1.95\textdegree\,HTG).}
Fan diagrams for contact pairs (\textbf{A}) R1-R2, (\textbf{B}) R2-R3, and (\textbf{C}) L2-R2. Grayed out portions of the data correspond to points in the fan where the measured current returning through the drain fell below 90\% of the applied value. 
(\textbf{D-F}) Linecuts along the $C=-6$ St\v{r}eda line for (\textbf{A}-\textbf{C}), respectively.
Black dashed line in (\textbf{F}) corresponds to $-h/6e^2$. 
Data are taken at $\SI{300}{mK}$.
}
\label{sup_fig:HM05}
\end{figure*}

We present two longitudinal measurements and one Hall measurement in Device 2 (Fig.~\ref{sup_fig:HM05}). 
All three contact pairs are consistent with a twist angle of $\sim\SI{1.95}{\degree}$. 
We observe a similar $C=-6$ state emanating from the CNP with the correct St\v{r}eda relation in this sample. 
The longitudinal resistance reaches a minimum value of $\SI{1.8}{k\Omega}$.
The Hall resistance is not well quantized, as one would expect for such a large longitudinal resistance. 
Quantization in this sample appears to be limited by device quality; we do not observe any evidence of Hofstadter physics as seen at similar temperatures and magnetic fields in the main device.

\begin{figure*}[h]
\centering
\includegraphics[width=7in]{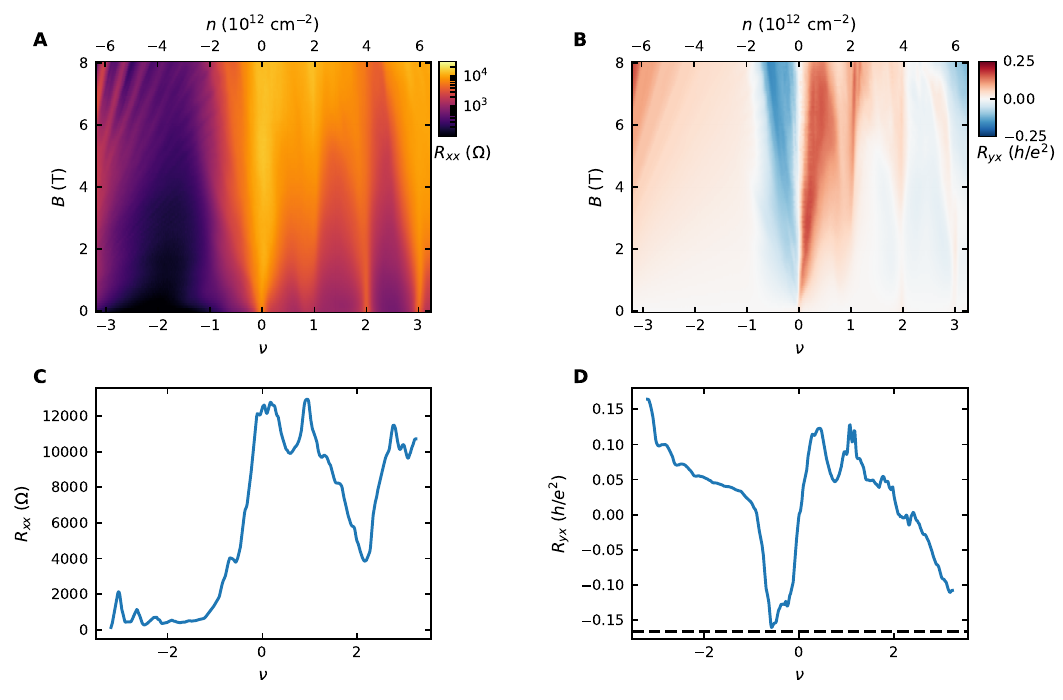}
\caption{\textbf{Landau Fan for Device 3 (1.85\textdegree\,HTG).}
Fan diagrams for (\textbf{A}) L1-L2 and (\textbf{B}) L1-R1.
Line cuts of the (\textbf{C}) longitudinal and (\textbf{D}) Hall resistance at $\SI{8}{T}$. 
Black dashed line corresponds to $-h/6e^2$. 
Data are taken at $D/\epsilon_0=\SI{0}{V/nm}$ and $\SI{495}{mK}$.
}
\label{sup_fig:HM06}
\end{figure*}

In Device 3, we see less clear signatures of a $C=-6$ state.
Device 3 is a $\sim\SI{1.85}{\degree}$ HTG device.
In the Landau fan, we see signatures of correlated states at integer fillings of the conduction flat band (Fig.~\ref{sup_fig:HM06}).
Comparing to the Landau fan at similar temperature for the main device (Fig.~\ref{sup_fig:maglabfan}), the CNP of device 3 is broader and lower in resistance, perhaps indicating more disorder in the device. 
Signatures of Landau levels emanating from $\nu=-4$ are much crisper, perhaps due to a larger bandwidth in the flat valence band due to the larger twist angle.
There is a very faint suppression of the longitudinal resistance emanating from the CNP roughly following the St\v{r}eda relation of a $C=-6$ state. 
Concurrently with this dip in longitudinal resistance, the Hall resistance is $R_{yx}=-0.16\,{h/e^2}$, near the value $-h/6e^2$ despite the large longitudinal resistance.

As discussed above, these two additional devices both exhibit signatures of a $C=-6$ state. 
Due to their differing twist angles, they exhibit varying strengths of zero-field correlated states at integer fillings of the flat conduction band; we do not see evidence of similar states for hole doping in any of the devices. 
We have shown that the supermoir\'e can be imaged when the HTG surface is exposed. 
Given the sensitivity of the supermoir\'e domain structure, it is possible that the precise structure will change during subsequent encapsulation, fabrication steps, and thermal cycling~\cite{hoke_imaging_2024}. 
The poor quantization and relatively large longitudinal resistance of Landau levels coming from $\nu=-4$ in all devices compared to typical high-mobility graphene devices suggests the presence of supermoir\'e domain walls in all samples.
As was argued for the AHE at odd integer fillings, the deviations from quantized Hall resistance and vanishing longitudinal resistance will depend sensitively on the structure of the domain wall network. 

% ---------------------------------------------------------------------------
% ---------------------------------------------------------------------------

% ---------------------------------------------------------------------------
% A: 43-44, 1.95
% B: 1-44, 1.96 degrees
% C: 5-44
% ---------------------------------------------------------------------------

% ---------------------------------------------------------------------------
\begin{figure*}[h]
\centering
\includegraphics{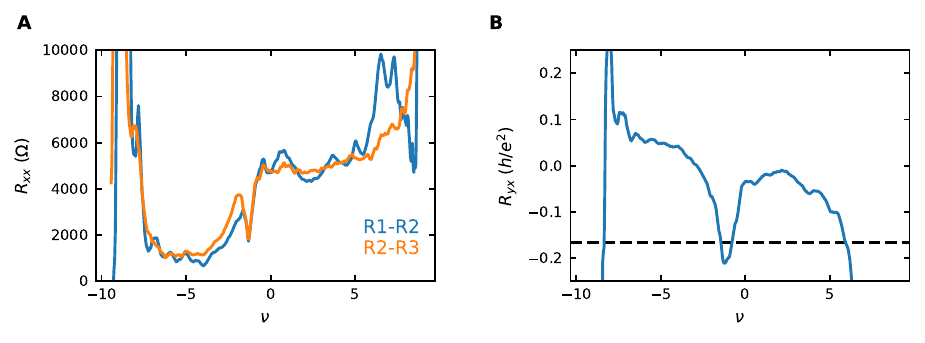}
\caption{\textbf{Line cuts of Landau Fan at 8 T for Device 2 (1.95\textdegree\,HTG).}
Line cuts of the (\textbf{A}) longitudinal and (\textbf{B}) Hall resistance of Fig.~\ref{sup_fig:HM05}. 
In (\textbf{A}), blue and orange correspond to R1-R2 and R2-R3, respectively. In (\textbf{B}), the dashed horizontal line corresponds to $-h/6e^2$.
}
\label{sup_fig:HM05_cut}
\end{figure*}
% ---------------------------------------------------------------------------

% ---------------------------------------------------------------------------
% ---------------------------------------------------------------------------

%%%%%%%%%%%%%%%%%%%%%%%%%%%%%%%%%%%%%%%%%%%%%%%%%%%%%%%%%
\subsection{Generalized BM Model and Momentum Dependent Tunneling}\label{sup_sec:BM}

In this section, we summarize the $B=0$ Bistritzer-MacDonald continuum model used to obtain the moir\'e band structures for the h-HTG and $\bar{\text{h}}$-HTG domains~\cite{devakul_magic-angle_2023}. Our presentation follows that of Refs.~\cite{xia_topological_2025, kwan2024fractionalchernmosaicsupermoire}, where more details of the modeling can be found. We first briefly discuss how a moir\'e continuum model arises from the supermoir\'e structure. We let the atomic lattice vectors for layer $l=1, 2, 3$ be the columns of $A_l= \lambda_{l}R(\theta_{l})A$, where $R(\theta)$ is the counter-clockwise rotation matrix, and $(\theta_1, \theta_2, \theta_3)=(\theta, 0, -\theta)$ are the twist angles. ${A} = a\begin{pmatrix}1 & \frac{1}{2} \\ 0 & \frac{\sqrt{3}}{2}\end{pmatrix}$, with $a=0.246\,$nm the graphene lattice constant, represents the untwisted and unstrained monolayer graphene lattice vectors. The dimensionless $(\lambda_1, \lambda_2, \lambda_3)=(1, 1/\cos(\theta), 1)$ parameterize the biaxial strain in the relaxed domains. These values are such that the moir\'e patterns between layers 1 and 2, and layers 2 and 3, become commensurate with each other, allowing for a moir\'e-periodic continuum model within a domain.

For valley $K$, the moir\'e continuum model takes the general form 
\begin{equation}\label{eq:nonint_BM}
H_K=\begin{bmatrix}
    -iv\bm{\sigma}_\theta\cdot\nabla+U\sigma_0 & T(\bm{r}-\bm{d}_t) & 0 \\
    T^\dagger(\bm{r}-\bm{d}_t) & -iv\bm{\sigma}\cdot\nabla & T(\bm{r}-\bm{d}_b)\\
    0 & T^\dagger(\bm{r}-\bm{d}_b) & -iv\bm{\sigma}_{-\theta}\cdot\nabla-U\sigma_0
\end{bmatrix},
\end{equation}
where the matrix acts in layer and sublattice $\sigma=A,B$ space, which is ordered according to $(1A,1B,2A,2B,3A,3B)$. We define $\bm{\sigma}=(\sigma_x,\sigma_y)$ which acts on sublattice space, and the monolayer Dirac velocity is taken to be $v=8.8\times 10^5\,\text{ms}^{-1}$. We also define the rotated Pauli matrices $\bm{\sigma}_{\theta}=e^{-i\theta\sigma_z}\bm{\sigma}$. $U$ is the interlayer potential used to model an external displacement field $D$. The relative interlayer moir\'e shifts for h-HTG satisfy $\bm{d}_t-\bm{d}_b=\frac{1}{3}(\bm{a}_2-\bm{a}_1)$, where $\bm{a}_{1,2}=\frac{4\pi}{3k_\theta}(\pm\frac{\sqrt{3}}{2},\frac{1}{2})$ are the basis moir\'e vectors in real-space, and $k_\theta=\frac{8\pi}{3a}\sin\frac{\theta}{2}$ is the moir\'e wavevector.
For $\bar{\text{h}}$-HTG, we take instead $\bm{d}_t-\bm{d}_b=-\frac{1}{3}(\bm{a}_2-\bm{a}_1)$. The Hamiltonian for valley $K'$ can be obtained by time-reversal symmetry.

The interlayer tunneling takes the form
\begin{equation}
\begin{gathered}
    T(\bm{r})=\begin{bmatrix}
        w_{AA}t_0(\bm{r}) & w_{AB}t_{-1}(\bm{r})\\
        w_{AB}t_1(\bm{r}) & w_{AA}t_0(\bm{r})
    \end{bmatrix}\\
    t_\alpha(\bm{r})=\sum_{n=0}^{2} e^{\frac{2\pi i}{3}n\alpha} e^{i\bm{q}_{n}\cdot\bm{r}}(1 -i \lambda_{\mathrm{MDT}} \hat{\bm{q}}_{n,\perp}\cdot \vec{\nabla}),
    % q_{n,x}+iq_{n,y}=-ik_\theta e^{\frac{2\pi i}{3}n},
\end{gathered}
\end{equation}
where $\bm{q}_n=k_\theta[\sin(\frac{2\pi}{3}n),\cos(\frac{2\pi}{3}n)]$, and $\hat{\bm{q}}_{n,\perp}= \left[\cos(\frac{2\pi}{3}n),\sin(\frac{2\pi}{3}n)\right]$ is a unit vector perpendicular to $\bm{q}_n$. We take $(w_{AA},w_{AB})=(77\,\text{meV},110\,\text{meV})$ for the intra- and inter-sublattice tunneling amplitudes. 

If $\lambda_\text{MDT}=0$, then $T(\bm{r})$ corresponds to standard local interlayer tunneling in the first harmonic approximation~\cite{bistritzer_moire_2011, devakul_magic-angle_2023}. The resulting continuum model is particle-hole symmetric (this is only weakly broken by the neglected Pauli twists in the kinetic terms). Accounting for the finite in-plane range of interlayer tunneling leads to momentum-dependent tunneling (MDT). It has been found~\cite{kwan2024fractionalchernmosaicsupermoire} that MDT can explain the substantial particle-hole asymmetry observed in the transport experiment of Ref.~\cite{xia_topological_2025}. In particular, correlated phenomena, including AHE, were found at $\nu>0$, but not at $\nu<0$. In this work, we take $\lambda_\text{MDT}\simeq -2.06\,$\r{A}.

%%%%%%%%%%%%%%%%%%%%%%%%%%%%%%%%%%%%%%%%%%%%%%%%%%%%%%%%%
\subsection{Orbital Zeeman Model}\label{sup_sec:gk_calcs}

% ---------------------------------------------------------------------------
\begin{figure*}[t]
\centering
\includegraphics[width=15cm]{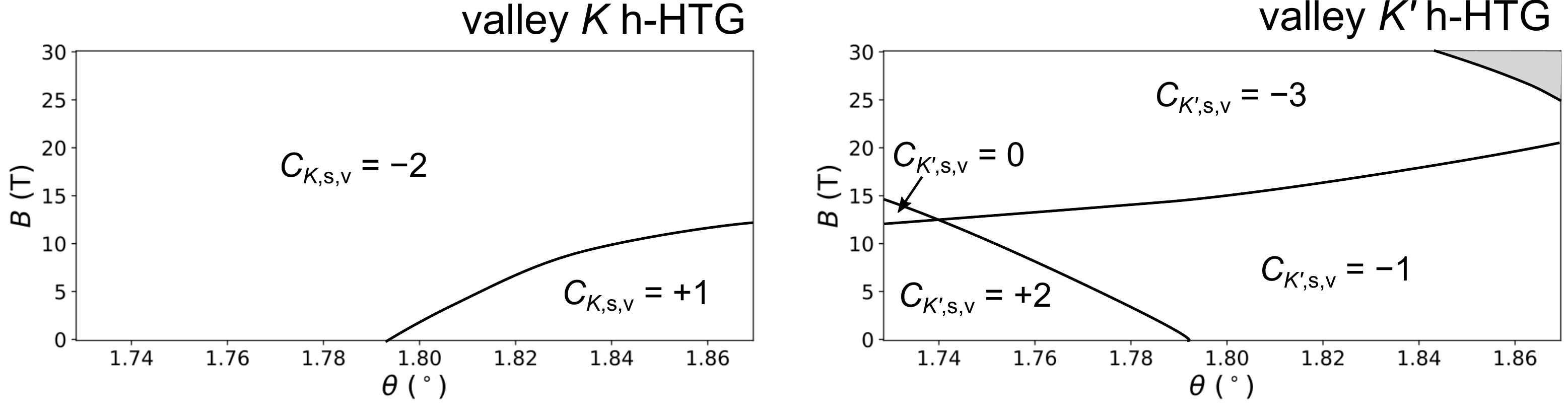}
\caption{\textbf{Chern numbers in the orbital Zeeman model.} $C_{\tau,s,\text{v}}$ indicates the Chern number of the central valence band in valley $\tau$ within one spin sector for h-HTG. For $K'$ h-HTG, the gray shaded region in the top-right contains tiny gaps and multiple transitions. The phase diagrams for $B<0$ and $\bar{\text{h}}$-HTG can be deduced using time-reversal and $C_{2z}$ operations. }
\label{sup_fig:gk_phase}
\end{figure*}
% ---------------------------------------------------------------------------

% ---------------------------------------------------------------------------
\begin{figure*}[t]
\centering
\includegraphics[width=17cm]{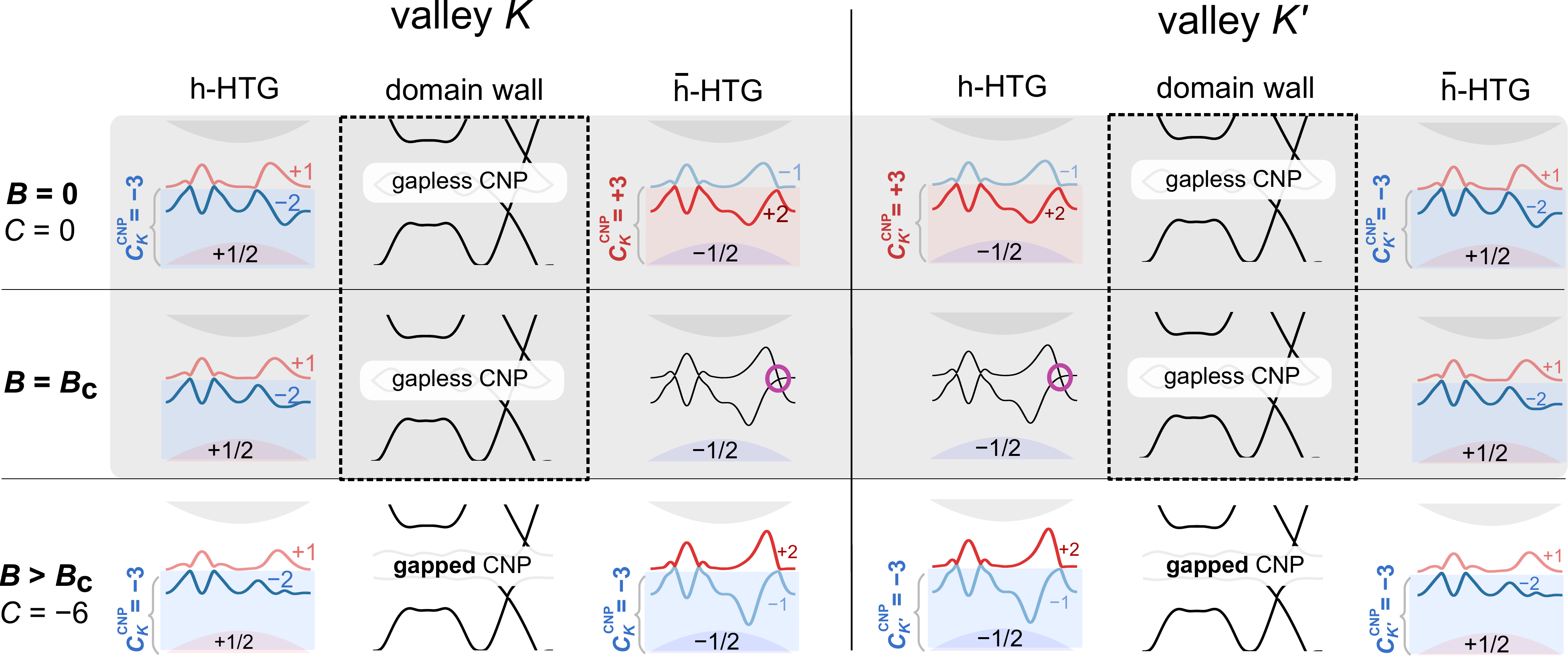}
\caption{\textbf{Schematic of globally gapped and gapless regimes in orbital Zeeman model.}
Schematic comparison of valley-resolved Chern numbers between h-HTG and $\bar{\text{h}}$-HTG domains for valley $K$ (left half) and $K'$ (right half). The central bands are labeled with their individual Chern numbers. We also indicate the half-integer values for the set of remote valence bands. $C_{2z}$ fixes the Chern numbers of one valley in h-HTG to be the same as those of the other valley in $\bar{\text{h}}$-HTG.
At $B=0$, the valley-resolved total Chern number at the CNP varies between h and $\bar{\text{h}}$ domains, forming gapless modes on the domain wall at the CNP.
At $B=B_c$, a topological transition within the domains (see magenta circles) occurs only in $K'$ h-HTG and $K$ $\bar{\text{h}}$-HTG, leading to $C=-6$ throughout the sample.
For $B>B_c$, the total valley-resolved Chern numbers $C^\text{CNP}_{K},C^\text{CNP}_{K'}$ of the CNP gap, which include a factor of two from spin, are identical for both domains, removing the topological protection of the edge modes and enabling a gapped domain wall at the CNP. Dispersions of the central bands are computed using the orbital Zeeman model at $\theta=1.77^\circ$ for $B=0.0\,\text{T},\ 5.9\,\text{T},\ 11.0\,\text{T}$, respectively.
Dispersion on the domain wall is for illustration purposes only, as the domain wall has moir\'e translational symmetry only in one direction.
}
\label{sup_fig:gk_schematic}
\end{figure*}
% ---------------------------------------------------------------------------

In this section, we discuss a model that approximates the influence of the external $B$-field via the orbital Zeeman effect. Throughout, we neglect spin Zeeman as it does not affect the topology of the bands. The gap between the central bands and the remote bands is large $\sim 100\,$meV. Therefore, following Refs.~\cite{sun2020topologicalzeeman, zhang_spin-polarized_2022}, we treat the remote bands perturbatively to extract an effective momentum-resolved orbital $g$-factor acting within the central band subspace
\begin{equation}
    g_{\tau,mn}(\bm{k})=\frac{ie}{4\mu_B\hbar}\sum_r \left(v^x_{\tau,mr}(\bm{k})v^y_{\tau,rn}(\bm{k})-v^y_{\tau,mr}(\bm{k})v^x_{\tau,rn}(\bm{k})\right)\left[\frac{1}{\epsilon_{\tau,m}(\bm{k})-\epsilon_{\tau,r}(\bm{k})}+\frac{1}{\epsilon_{\tau,n}(\bm{k})-\epsilon_{\tau,r}(\bm{k})}\right],
\end{equation}
where the velocity operator is $v^j_{ab}(\bm{k})=\bra{u_{\tau,a}(\bm{k})}\frac{\partial H_\tau(\bm{k})}{\partial k_j}\ket{u_{\tau,b}(\bm{k})}$, $\mu_B=\frac{e\hbar}{2m_e}$ is the Bohr magneton, $\tau$ is the valley index, $m,n$ are central band indices, and $r$ is a remote band index. 

At finite $B$, we model the Hamiltonian of the central bands as
\begin{equation}
    [H^{\text{eff},B}_\tau(\bm{k})]_{mn}=\delta_{m,n}\epsilon_{\tau,m}(\bm{k})-g_{\tau,mn}(\bm{k})\mu_B B,
\end{equation}
where $\epsilon_{\tau,m}(\bm{k})$ is the band dispersion at $B=0$. By solving $H^{\text{eff},B}_\tau(\bm{k})$, we obtain the effective dispersions and Chern numbers of the central bands at finite $B$. Note that this framework preserves the moir\'e translation symmetry within a domain, and does not capture any physics of the Landau levels or the Hofstadter spectrum. Hence, this method is only justified for small $B$ before such physics becomes important, though we perform calculations with the orbital Zeeman model for larger $B$ for completeness. We do not incorporate interaction effects, which may renormalize the positions of the phase boundaries. 
Note that at the semi-classical level, the magnetic field also changes the density carried by a Chern band. This leads to the St\v{r}eda anomaly, and is not directly captured by $g_{\tau,mn}(\bm{k})$. 

Within a domain, time-reversal relates valley $\tau$ at magnetic field $B$ to valley $\bar{\tau}$ at magnetic field $-B$, where $\bar{\tau}$ indicates the opposite valley to $\tau$. $C_{2z}$ relates valley $\tau$ in h-HTG to valley $\bar{\tau}$ in $\bar{\text{h}}$-HTG at the same magnetic field. Hence, we can present the results just for $B\geq0$ in h-HTG, as shown in Fig.~\ref{sup_fig:gk_phase}. At $B=0$ where there is time-reversal symmetry, a critical angle $\theta\sim 1.8^\circ$ separates a regime where $C_{
K,\text{v}}=-2$ from another regime where $C_{K,\text{v}}=+1$. For $B>0$, the $C_{K,s,\text{v}}=+1$ region transitions to $C_{K,s,\text{v}}=-2$ via three gap closings along the $\gamma-m$ lines in the mBZ. The critical field $B_c$ vanishes at the zero-field critical twist angle dividing the two regimes. For valley $K'$, there is a similar transition for $\theta\lesssim 1.8^\circ$ that lowers $C_{K',s,\text{v}}$ by 3. However, we find another set of gap closings at the $\kappa,\kappa'$ corners that reduces $C_{K',s,\text{v}}$ by 2. We note though that the latter transition occurs at large $B\gtrsim 10\,$T where the orbital Zeeman model may be unreliable.

As discussed in the main text, the valley Chern number of the gap above the central valence band inherits additional contributions owing to the remote valence bands~\cite{nakatsuji2023multiscale, kwan2024strong}. This can be accounted for by taking $C_{K,s,\text{remote v}}=+1/2$ and $C_{K',s,\text{remote v}}=-1/2$. Then, the total valley-resolved Chern number in the CNP gap above the central valence band is $C^\text{CNP}_{\tau}=2(C_{\tau,s,\text{remote v}}+C_{\tau,s,\text{v}})$, where the factor of two accounts for the spin degree of freedom (the spin Zeeman effect does not affect the topology of the effective bands). This is schematically illustrated in Fig.~\ref{sup_fig:gk_schematic}.

\end{document}